\documentclass{ws-rv9x6}
\usepackage{subfigure}   
\usepackage{ws-rv-van}   
\makeindex

\usepackage{graphicx}

\begin{document}

\chapter[Comparison of EoS models with different cluster dissolution mechanisms]{Comparison of equation of state models with different cluster dissolution mechanisms}\label{ra_ch1}

\author{Helena Pais}
\address{CFisUC, Department of Physics, University of Coimbra,\\
P-3004-516 Coimbra, Portugal, \\
pais.lena@uc.pt}
\author[H. Pais and S. Typel]{Stefan Typel}
\address{Institut f\"{u}r Kernphysik, Technische Universit\"{a}t Darmstadt,\\
Schlossgartenstra\ss{}e 9, D-64289 Darmstadt, Germany\\
GSI Helmholtzzentrum f\"{u}r Schwerionenforschung GmbH,\\
Planckstra\ss{}e 1, D-64291 Darmstadt, Germany,\\
s.typel@gsi.de}

\begin{abstract}
The appearance of nuclear clusters in stellar matter at densities 
below nuclear saturation is an important feature in the modeling of the 
equation of state for astrophysical applications. 
There are different theoretical concepts to describe 
the dissolution of nuclei with increasing density and temperature.
In this contribution, the predictions of two approaches are compared: the medium dependent
change of the nuclear masses in a generalized relativistic density functional approach and
the excluded-volume mechanism in a statistical model. Both approaches use
the same description for the interaction between the nucleons. The composition of 
neutron star matter, in particular the occurrence of light and heavy nuclei,
and its thermodynamic properties are studied.
\end{abstract}
\body


\section{Introduction} 
\label{sec:intro}

In astrophysical simulations of core-collapse supernovae 
(CCSN)\cite{Mezzacappa:2005ju,Janka:2006fh,Janka:2012wk,Burrows:2012ew}
and the description of compact star 
properties\cite{Glendenning:1997wn,Weber:2004kj,Haensel:2007yy,Potekhin:2011xe,Lattimer:2012nd,Hebeler:2013nza}, 
the equation of state (EoS) is an essential ingredient. It provides the information on 
thermodynamic properties of strongly interacting 
matter with baryons and leptons as basic degrees of freedom
\cite{Lattimer:1991nc,Shen:1998by,Shen:1998gq,Klahn:2006ir,Lattimer:2006xb,Hempel:2009mc,Shen:2011qu,Shen:2011kr,Shen:2011fc,Steiner:2012rk}. 
The description of such matter at densities below the 
nuclear saturation density $n_{\rm sat} \approx 0.16$~fm$^{-3}$
and temperatures $T \lesssim 15$~MeV represents a particular challenge for 
theoretical models, where the short-range strong interaction 
and the long-range electromagnetic interaction have to be considered explicitly.
The competition of these forces with the entropy leads to the formation 
and dissolution of inhomogeneous structures on mesoscopic length scales 
with different geometries. This is in contrast to the liquid-gas phase 
transition in pure nuclear matter 
where the Coulomb interaction and 
the charge of particles are neglected, see, e.g., Refs.\cite{Barranco:1980zz,Muller:1995ji,Bugaev:2000ar,Gulminelli:2003bd,Ducoin:2005aa,Ducoin:2006jd,Ducoin:2006td,Hempel:2011mk,Furusawa:2011wh,Hempel:2013tfa,Furusawa:2013rta,Fischer:2013eka} for details. 
In stellar matter the 
phenomenon of frustration is observed since the available space of 
thermodynamic variables is reduced due to the specific condition 
of charge neutrality.

In dilute matter 
at temperatures above $2$~MeV and with densities up to approximately $0.001~n_{\rm sat}$, 
light nuclei (deuterons, tritons, helions, $\alpha$-particles) 
are the most relevant nuclear species that are formed as 
many-nucleon correlations \cite{Typel:2009sy,Horowitz:2005nd,Ducoin:2006td,Heckel:2009br, 
Ferreira:2012ha,Avancini:2010xb,Avancini:2012bj}. With increasing densities 
and at lower temperatures,
also heavier nuclei with increasing mass numbers appear and 
the chemical composition of matter changes.
Finally, when the saturation density $n_{\rm sat}$ is approached, 
the occurrence of so-called ''pasta phases'' with several 
different geometries is 
expected\cite{Ravenhall:1983uh,Williams:1985,Horowitz:2004pv,Horowitz:2005zb,Maruyama:2005vb,Watanabe:2004tr,Sonoda:2007sx,Pais:2012js,Grill:2014aea}, before the system converts to uniform matter composed of nucleons, electrons and muons.

The complex structure of stellar matter 
is particularly important in astrophysical applications. 
The formation of clusters and pasta phases in CCSN matter 
affects the neutrino opacity \cite{Horowitz:2004yf,Horowitz:2004pv,Horowitz:2005zb,Sonoda:2007ni,Alloy:2010fk}.
This quantity plays a central role in the energy transport and deposition. 
Thus it can affect the development of a shock wave during the 
supernova collapse \cite{Williams:1985,Furusawa:2013tta}
and the cooling of the proto neutron star 
(PNS)\cite{Haensel:1994kp,Page:2013uwa}. 
The crust cooling will also affect its thickness, and, consequently, the moment of inertia, which will have direct influence on the interpretation of pulsar glitches\cite{Haskell:2015jra}. 
The composition of matter 
also determines the structure of the crust of cold catalyzed neutron stars
where the additional condition of $\beta$-equilibrium fixes the 
isospin asymmetry of the nuclear subsystem\cite{Salpeter:1961zz,Baym:1971pw,Haensel:1993zw,Douchin:2001sv,Ruester:2005fm,Pearson:2011zz,Wolf:2013ge}.

Statistical models with nucleons, nuclei and charged leptons as 
degrees of freedom are a popular approach to describe stellar matter 
at subsaturation densities\cite{Botvina:2003dm,Botvina:2008su,Hempel:2009mc,Blinnikov:2009vm,Raduta:2009tp,Raduta:2010ym,Gulminelli:2011hr,Hempel:2011mk,Buyukcizmeci:2012it,Buyukcizmeci:2013bza,Raduta:2013fna,Sagun:2013moa}. The interaction between the constituents
is often incorporated on the basis of nonrelativistic Skyrme or 
relativistic density functionals, employing a mean-field picture.
In most cases, only the strong interaction between the nucleons is 
considered in this way.
In order to model the transition from clustered matter to uniform matter 
at high densities and temperatures, the dissolution of nuclei 
has to be implemented in the theoretical approach. Various prescriptions
are employed to suppress the occurrence of clusters.

A widely used heuristic approach is the excluded-volume mechanism 
using a geometrical picture\cite{Hempel:2009mc,Hempel:2011mk,Hempel:2011kh,Typel:2016srf}. 
A finite volume, usually proportional 
to the mass number $A$, is attributed to each nucleon or nucleus,
leaving a reduced volume for the motion of the remaining particles.
As a consequence, the system lowers the fraction of larger-size
constituents in favor of smaller-size particles
with increasing density. Hence, nuclei are converted to nucleons.
The effect of the finite particle volume on the system properties 
can be interpreted as the action of an effective repulsive, 
hard-core interaction. The excluded-volume approach is implemented, e.g.,
in an EoS model that considers an ensemble of nucleons, nuclei and electrons
in statistical equilibrium \cite{Hempel:2009mc}. The interaction between the nucleons is
described in a relativistic mean-field (RMF) model.
For  this  model,  EOS  tables for a variety of 
RMF parametrizations \cite{Hempel:2009mc,Hempel:2011mk,Steiner:2012rk,Fischer:2013eka} 
are available online \cite{compose}. They include 
the full distribution of light and heavy  nuclei and  cover a  
broad range in density, temperature, and isospin asymmetry.

In a more microscopic description, nuclei will dissolve with 
increasing density, mainly as a result of the Pauli principle \cite{Sumiyoshi:2008qv,Typel:2009sy,Ropke:2011tr,Hempel:2011kh,Ropke:2012qv,Ropke:2014uma,Ropke:2014fia}. 
At high densities, it becomes difficult to form clusters as 
many-nucleon correlations, since most of the momentum space is already occupied
by background nucleons. The essential consequence is a reduction of the 
cluster binding energy with increasing density. This effect becomes smaller, however,
at higher temperatures due to a more dilute population of states in 
momentum space. Such medium-dependent mass shifts can be extracted 
from microscopic calculations of few-body correlations in dense matter.
They are used in a parametrized form in density functionals for stellar matter.
A reduction of the binding energy causes a lowering of the particle abundance
simply because of statistical reasons.
The mass-shift method is realized in the generalized relativistic 
density functional (gRDF) approach \cite{Typel:2009sy,Voskresenskaya:2012np,Typel:2015kga},
which is an extension of a conventional RMF model for nuclear matter
with density dependent (DD) couplings \cite{Typel:1999yq}.


In this work, we compare the formation and dissolution of light and 
heavy clusters in dilute matter within these two models: 
the statistical model with excluded-volume mechanism \cite{Hempel:2009mc},
denoted as HS model in the following,
and the gRDF approach with 
cluster mass shifts\cite{Typel:2009sy,Voskresenskaya:2012np,Typel:2015kga}.
For this comparison, it is important to note that the same 
parametrization (DD2) of the RMF interaction is employed in both models
such that differences in the predictions can be attributed to the cluster description. 
Besides the density dependence of the cluster suppression, we will 
investigate the predictions of the two models as a function 
of the temperature. Here, different procedures to include thermal 
excitations of the nuclei, which depend on the employed level densities, 
become relevant.

The paper is organized as follows. In section \ref{sec:theo}, the 
theoretical formulation of the excluded-volume mechanism and its particular
realization in the HS model and the gRDF approach are discussed.
Predictions of the HS and gRDF models for neutron star matter, i.e.\ stellar matter
in $\beta$-equilibrium, are presented in section \ref{sec:res}. The main emphasis
is on the comparison of the chemical composition, in particular the occurrence
of light and heavy nuclei. Also thermodynamic properties are briefly compared.
Conclusions are given in section \ref{sec:concl}.

\section{Theoretical formalism}
\label{sec:theo}

All thermodynamic properties of stellar matter can be determined once
a suitable thermodynamic potential is specified. In most cases, a 
Helmholtz free energy density $f(T,\{n_{i}\})$ 
or a grand canonical potential density $\omega(T,\{\mu_{i}\})$ is given.
They depend on the temperature $T$ and the number densities $n_{i}$ 
or chemical potentials $\mu_{i}$ of all constituents $i$. 
In both cases, the excluded-volume mechanism and 
the medium-dependence of mass shifts can be incorporated in the formalism. However,
one has to make sure that the theory is thermodynamically consistent, 
i.e.\ the usual definitions of thermodynamic quantities and relations for 
derivatives hold after the modification
of the original thermodynamic potential.

A general formulation of the excluded-volume mechanism and mass shifts
is presented in the following
subsection \ref{sec:exvol},  
assuming relativistic kinematics of the particles and general 
particle statistics. This model incorporates effects from the 
interaction between particles in the framework of a RMF approach
with density dependent couplings. In subsection \ref{sec:HS}, 
the relation of this
formulation to the HS model is established and differences are explored.
The formalism of the gRDF approach is presented in subsection \ref{sec:gRDF}.

\subsection{RMF model with excluded-volume mechanism and mass/energy shifts}
\label{sec:exvol}


The excluded-volume mechanism is a simple means to model a repulsive interaction 
between particles. A very general theoretical formulation was presented 
in Ref.\cite{Typel:2016srf} that is adapted to the present application. 
Every particle is assumed
to have a finite volume 
that reduces the total volume $V$ of the system to the available volume $V_{i}$ for the motion 
of a particle $i$. Hence, this available volume is
\begin{equation}
 V_{i} = V \Phi_{i}
\end{equation}
with the available volume fraction $\Phi_{i}$. In the following
it is supposed that the functions $\Phi_{i}$ only depend on the  
number densities $n_{j}=N_{j}/V$ of all particles $j$ in the system.
The geometric picture of rigid spheres 
in the excluded-volume mechanism corresponds
to the choice
\begin{equation}
\label{eq:phi_i}
 \Phi_{i} = 1 - \sum_{j} v_{ij} n_{j}
\end{equation}
with coefficients $v_{ij}$ that are connected to the radii $R_{i}$ 
of the particles as
\begin{equation}
 v_{ij} = \frac{2\pi}{3} \left( R_{i} + R_{j} \right)^{3}  \: .
\end{equation}
This formulation is symmetric in the indices $i$ and $j$ as required
by the consistency with the virial equation of state at low densities
for particles with different radii \cite{Typel:2016srf}. 
Other functional dependencies than (\ref{eq:phi_i}) are permissible in general.
However, the rigid sphere picture does not apply any more in this case.
The particle number densities $n_{i} = n_{i}^{(v)}$ are the vector densities 
in the relativistic description. They have to be distinguished from the scalar
densities $n_{i}^{(s)}$. Both will be defined below.

The total grand canonical potential density of the system can be written as 
\begin{equation}
\label{eq:omega}
 \omega(T,\{\mu_{i}\}) = \sum_{i} \omega_{i} + \omega_{\rm meson}
 - \omega^{(r)}
\end{equation}
with three distinct terms. 
The single quasi-particle contribution
\begin{equation}
\label{eq:omega_i}
 \omega_{i} = \Phi_{i} \omega_{i}^{(0)} 
\end{equation}
contains the available volume fraction $\Phi_{i}$ and the
standard expression
\begin{equation}
\label{eq:omega_i_0}
 \omega_{i}^{(0)}(T,m_{i}^{\ast},\mu_{i}^{\ast}) = - T \frac{g_{i}}{\sigma_{i}} 
  \int \frac{d^{3}k}{(2\pi)^{3}} \:
 \ln \left\{ 1 + \sigma_{i} \exp\left[ -\frac{E_{i}(k,m_{i}^{\ast})
 -\mu_{i}^{\ast}}{T}\right]\right\}
\end{equation}
with the particle degeneracy factor $g_{i}$, the effective mass
\begin{equation}
\label{eq:m_eff}
 m_{i}^{\ast} = m_{i} - S_{i} \: ,
\end{equation} 
and the effective chemical potential
\begin{equation}
\label{eq:mu_eff}
 \mu_{i}^{\ast} = \mu_{i} - V_{i} \: .
\end{equation}
The appearance of the scalar potential $S_{i}$ and the vector potential $V_{i}$ are typical
for the RMF approach.
The quantity $\sigma_{i}$ encodes the particle statistics.
The case $\sigma_{i}=+1$ corresponds to Fermi-Dirac particles 
and $\sigma_{i}=-1$ to Bose-Einstein particles. In the limit $\sigma_{i} \to 0$, 
the result for Maxwell-Boltzmann statistics 
\begin{eqnarray}
\label{eq:omega_i_MB}
 \omega_{i}^{(0)}(T,m_{i}^{\ast},\mu_{i}^{\ast}) & = & - T g_{i}
  \int \frac{d^{3}k}{(2\pi)^{3}} \:
 \exp\left[ -\frac{E_{i}(k,m_{i}^{\ast})-\mu_{i}^{\ast}}{T}\right]
\end{eqnarray}
is recovered.
Effects of the finite particle volumes are taken into account by the 
density dependent prefactor $\Phi_{i}$ in (\ref{eq:omega_i}).
For alternative formulations of the excluded-volume mechanism see 
Ref.\cite{Typel:2016srf}. 

The relativistic relation
\begin{equation}
  \label{eq:disp}
  E_{i}(k,m_{i}^{\ast}) = \sqrt{k^{2} + \left( m_{i}^{\ast} \right)^{2}} 
\end{equation}
connects the momentum $k$ with the energy $E_{i}$ of each particle, which
is considered as a quasi-particle
since scalar and vector potentials $S_{i}$ and $V_{i}$ 
appear in the definition of the effective mass (\ref{eq:m_eff}) and the
effective chemical potential (\ref{eq:mu_eff}). These potentials are given by
\begin{equation}
\label{eq:S}
 S_{i} = C_{\sigma} g_{i\sigma} n_{\sigma} + C_{\delta} g_{i\delta} n_{\delta} 
 - \Delta m_{i}
\end{equation}
and 
\begin{equation}
\label{eq:V}
 V_{i} = C_{\omega} g_{i\omega} n_{\omega} + C_{\rho} g_{i\rho} n_{\rho} 
 + \Delta E_{i}
 + D_{i} V_{\rm meson}^{(r)} + V_{i}^{(r)} + W_{i}^{(r)}
\end{equation}
with contributions due to the coupling of the particles to meson fields, 
density dependent mass shifts $\Delta m_{i}$, energy shifts $\Delta E_{i}$
and rearrangement terms $V_{\rm meson}^{(r)}$, $V_{i}^{(r)}$, and $W_{i}^{(r)}$. The factors
\begin{equation}
\label{eq:C_j}
 C_{j}= \frac{\Gamma_{j}^{2}}{m_{j}^{2}}
\end{equation}
are found from the meson-nucleon couplings $\Gamma_{j}$ and masses $m_{j}$ of the considered
mesons $j=\omega, \sigma, \rho, \delta$ as usual in RMF models.
The couplings $\Gamma_{j}$ themselves depend 
on a number density 
\begin{equation} 
\label{eq:ntilde}
 \tilde{n} = \sum_{i} D_{i} n_{i}^{(v)}
\end{equation}
with factors $D_{i}$ to be specified below.
The source densities in equations (\ref{eq:S}) and (\ref{eq:V}) 
are given by 
\begin{equation}
\label{eq:source_v}
 n_{j} = \sum_{i} g_{ij} n_{i}^{(v)} 
\end{equation}
for the Lorentz vector mesons $j=\omega,\rho$ 
and 
\begin{equation}
\label{eq:source_s}
 n_{j} = \sum_{i} g_{ij} n_{i}^{(s)} 
\end{equation}
for the scalar mesons $j=\sigma,\delta$
with appropriate factors $g_{ij}$ and vector and scalar densities
$n_{i}^{(v)}$ and $n_{i}^{(s)}$, respectively, that are defined below.

The meson rearrangement potential
\begin{equation}
 V_{\rm meson}^{(r)} = \frac{1}{2} \left( C_{\omega}^{\prime} n_{\omega}^{2}
 + C_{\rho}^{\prime} n_{\rho}^{2}
 - C_{\sigma}^{\prime} n_{\sigma}^{2}
 - C_{\delta}^{\prime} n_{\delta}^{2} \right) \: ,
\end{equation}
in eq.\ (\ref{eq:V}) with
\begin{equation}
 C_{j}^{\prime} = \frac{dC_{j}}{d\tilde{n}}
\end{equation}
and the excluded-volume rearrangement term
\begin{equation}
 V_{i}^{(r)}  =   \sum_{j} \omega_{j} 
 \frac{\partial \ln \Phi_{j}}{\partial n_{i}^{(v)}}
\end{equation} 
also appear in the contribution
\begin{equation}
\label{eq:omega_r}
 \omega^{(r)} = \omega^{(r)}_{\rm meson} + \omega^{(r)}_{\Phi} + \omega^{(r)}_{\Delta}
\end{equation}
to the total grand canonical density (\ref{eq:omega}) with 
\begin{equation}
\label{eq:omega_meson_r}
 \omega^{(r)}_{\rm meson} = V_{\rm meson}^{(r)} \tilde{n}
\end{equation} 
and
\begin{equation}
\label{eq:omega_phi_r}
 \omega_{\Phi}^{(r)} = \sum_{i}  n_{i}^{(v)} V_{i}^{(r)} \: .
\end{equation}
The rearrangement term 
\begin{equation}
 W_{i}^{(r)}  =   \sum_{j} \left( n_{j}^{(s)}
 \frac{\partial \Delta m_{j}}{\partial n_{i}^{(v)}}
 + n_{j}^{(v)}
 \frac{\partial \Delta E_{j}}{\partial n_{i}^{(v)}}
 \right)
\end{equation} 
in the vector potential (\ref{eq:V}), which is
caused by the density dependence of the mass shifts and the energy shifts,
is contained in the contribution
\begin{equation}
\label{eq:omega_dm_r}
 \omega_{\Delta}^{(r)} = \sum_{i}  n_{i}^{(v)}  W_{i}^{(r)} 
\end{equation}
to (\ref{eq:omega_r}).
Finally, the meson term in (\ref{eq:omega})
is given by
\begin{equation}
\label{eq:omega_meson}
 \omega_{\rm meson} = -\frac{1}{2} \left( C_{\omega} n_{\omega}^{2}
 + C_{\rho} n_{\rho}^{2}
 - C_{\sigma} n_{\sigma}^{2}
 - C_{\delta} n_{\delta}^{2} \right) \: .
\end{equation}
The appearance of the rearrangement contributions guarantees the thermodynamic
consistency of the model. In particular, the particle number densities
\begin{equation}
\label{eq:n_i_v}
 n_{i} = - \left. \frac{\partial \omega}{\partial \mu_{i}} \right|_{T,\{ \mu_{j\neq i}\}}
 = n_{i}^{(v)} = g_{i} \Phi_{i} \int \frac{d^{3}k}{(2\pi)^{3}} \: f_{i}
\end{equation}
and the scalar densities
\begin{equation}
\label{eq:n_i_s}
 n_{i}^{(s)} = 
  \left. \frac{\partial \omega}{\partial m_{i}} \right|_{T,\{ \mu_{j}\}}
= g_{i} \Phi_{i} \int \frac{d^{3}k}{(2\pi)^{3}} \: f_{i}
 \frac{m_{i}^{\ast}}{\sqrt{k^{2}+\left(m_{i}^{\ast} \right)^{2}}}
\end{equation}
with the distribution function
\begin{equation}
 f_{i}(T,k,m_{i}^{\ast},\mu_{i}^{\ast}) 
 = \left\{ \exp \left[ \frac{E_{i}(k,m_{i}^{\ast})-\mu_{i}^{\ast}}{T}\right] + \sigma_{i}\right\}^{-1}
\end{equation}
assume the usual form for quasi-particles, however, including the available
volume fractions $\Phi_{i}$ as a prefactor. Because the functions $\Phi_{i}$ 
do not depend on the scalar densities in the models considered in the present paper, there are no rearrangement contributions in the scalar potential
(\ref{eq:S}). For the case of a dependence on the scalar densities, 
see Ref.\cite{Typel:2016srf}.

If the relativistic energy (\ref{eq:disp}) is replaced by the nonrelativistic approximation
\begin{equation}
 E_{i}(k,m_{i}^{\ast}) = m_{i}^{\ast}  + \frac{k^{2}}{2m_{i}^{\ast}} \: ,
\end{equation}
the number density (\ref{eq:n_i_v}) and
the single quasi-particle grand canonical potential density 
in Maxwell-Boltzmann statistics (\ref{eq:omega_i_MB}) can be given analytically
with the ideal gas results
\begin{equation}
\label{eq:n_i_MB}
 n_{i} = n_{i}^{(v)} = 
 \frac{g_{i}\Phi_{i}}{\left(\lambda_{i}^{\ast}\right)^{3}} 
 \exp \left( \frac{\mu_{i}^{\ast}-m_{i}^{\ast}}{T}\right)
\end{equation}
and 
\begin{equation}
\label{eq:omega_i_MB_nr}
 \omega_{i}(T,m_{i}^{\ast},\mu_{i}^{\ast}) = 
 -T n_{i}^{(v)}
\end{equation}
that contain the effective thermal wavelength
\begin{equation}
 \lambda_{i}^{\ast} =  \lambda(m_{i}^{\ast},T) = \sqrt{\frac{2\pi}{m_{i}^{\ast}T}}
\end{equation}
depending on the temperature and the effective mass.

For massless particles, $\omega_{i}^{(0)}$ can also be calculated analytically, e.g.,
as
\begin{equation}
\omega_{\gamma}^{(0)} = - g_{\gamma} \frac{\pi^{2}}{90} T^{4}
\end{equation}
for photons with degeneracy factor $g_{\gamma} = 2$.

Irrespective of the particle statistics and the dispersion relation for $E_{i}(k,m_{i}^{\ast})$,
the grand canonical potential density (\ref{eq:omega}) can be written as
\begin{equation}
\label{eq:omega_b}
 \omega(T,\{ \mu_{i}\}) = \sum_{i} \omega_{i} + \omega_{\rm meson}
 - \omega_{\rm meson}^{(r)} - \omega_{\Phi}^{(r)} - \omega_{\Delta}^{(r)} 
\end{equation}
taking the decomposition (\ref{eq:omega_r}) into account. 
The first and the next-to-last term can be combined as
\begin{eqnarray}
 \sum_{i} \omega_{i} - \omega_{\Phi}^{(r)}
 & = & \sum_{i} \left[ \omega_{i} - n_{i}^{(v)} V_{i}^{(r)} \right]
 = 
 \sum_{j}  \omega_{j} \sum_{i} \left[ \delta_{ij}  - n_{i}^{(v)} 
 \frac{\partial \ln \Phi_{j}}{\partial n_{i}^{(v)}} \right] \: .
\end{eqnarray}
With the choice (\ref{eq:phi_i}) for the available volume fractions $\Phi_{i}$, one finds
in particular
\begin{eqnarray}
 \sum_{i} \omega_{i} - \omega_{\Phi}^{(r)}
  & = & 
 \sum_{j} \Phi_{j} \omega_{j}^{(0)} \sum_{i} \left[ \delta_{ij}  + n_{i}^{(v)} \frac{v_{ji}}{\Phi_{j}}
  \right]
  \\ \nonumber & = & 
 \sum_{j}  \omega_{j}^{(0)} \left[ \Phi_{j} +  \sum_{i} n_{i}^{(v)} v_{ji}
  \right]
 = 
 \sum_{j}  \omega_{j}^{(0)} 
\end{eqnarray}
and thus
\begin{equation}
\label{eq:omega_simple}
 \omega(T,\{ \mu_{i}\}) = \sum_{i} \omega_{i}^{(0)} + \omega_{\rm meson}
 - \omega_{\rm meson}^{(r)} - \omega_{\Delta}^{(r)} 
\end{equation}
with the original single quasi-particle contributions (\ref{eq:omega_i_0}) without a prefactor
and no explicit
rearrangement term from the excluded-volume mechanism. 
Nevertheless, the dependence on the available
volume fractions $\Phi_{i}$ enters through the rearrangement term $V_{i}^{(r)}$ 
in the vector potentials $V_{i}$.

The entropy density of the system is obtained by the standard thermodynamic derivative 
\begin{eqnarray}
 \lefteqn{s  =  - \left. \frac{\partial \omega}{\partial T} \right|_{\{ \mu_{i}\}}}
 \\ \nonumber &  = &
 - \sum_{i} g_{i} \Phi_{i} \int \frac{d^{3}k}{(2\pi)^{3}} \:
 \left[ f_{i} \ln f_{i} + \frac{1-\sigma_{i}f_{i}}{\sigma_{i}} 
 \ln \left( 1- \sigma_{i} f_{i}\right)\right]
 - \sum_{i} \Phi_{i} \frac{\omega_{i}^{(0)}}{g_{i}} \frac{\partial g_{i}}{\partial T}
\end{eqnarray}
with the usual contribution and a term that takes an explicit temperature dependence of the
degeneracy factors $g_{i}$ into account, e.g., for nuclei, see below.

\subsection{gRDF model}
\label{sec:gRDF}

In this extension of a RMF model with density dependent couplings, 
nucleons, nuclei, electrons, muons and photons
are considered as the relevant particle species. In addition, two-nucleon correlations in 
the continuum are included as effective quasi-particles like nuclei 
with temperature dependent resonance
energies or masses. Nucleons, electrons, and muons are treated as fermions including their
antiparticles. Photons are added as massless bosons as usual. The correct Bose-Einstein or 
Fermi-Dirac statistics are taken into account for light nuclei 
(${}^{2}$H = 'd', ${}^{3}$H = 't', ${}^{3}$He = 'h', and ${}^{4}$He = '$\alpha$') 
and two-nucleon correlations in the ${}^{3}S_{1}$(np) and ${}^{1}S_{0}$(np, nn, and pp)
channels. Heavy nuclei with mass number $A>4$ are
described with nonrelativistic kinematics assuming Maxwell-Boltzmann statistics. Experimental
rest masses $m_{i}$ are used as far as available. For nuclei they are taken from the
2012 Atomic Mass Evaluation\cite{Wang:2012} if available, 
or they are calculated in the DZ10 model\cite{Duflo:1995ep}. All nuclei
within the neutron and proton driplines are included in the calculations. The driplines
were determined from the neutron and proton separation energies after removing 
the Coulomb contribution
of a homogeneously charge sphere of radius $r_{C} = 1.25 \cdot A^{1/3}$~fm
to the total binding energies. 

Since chemical equilibrium is assumed for the full ensemble,
the chemical potential of every particle is given by
\begin{equation}
\label{eq:mu_i}
 \mu_{i} = B_{i} \mu_{b} + Q_{i} \mu_{q} + L_{i}^{(e)} \mu_{l^{(e)}} + L_{i}^{(\mu)} \mu_{l^{(\mu)}}
\end{equation}
with baryon numbers $B_{i}$, charge numbers $Q_{i}$, electron lepton numbers $L_{i}^{(e)}$,
and muon lepton numbers $L_{i}^{(\mu)}$ and corresponding independent chemical potentials
of the conserved currents. The quantum numbers are specified in table \ref{tab:qn}
for all elementary particles of the model and nuclei $i=(A,Z)$ with 
mass numbers $A$, charge numbers $Z$, and neutron numbers $N=A-Z$. 
The chemical potential of an 
antiparticle is the negative of that of a particle. The lepton chemical potentials
are taken to be equal, i.e.,
\begin{equation}
 \mu_{l} = \mu_{l^{(e)}} = \mu_{l^{(\mu)}} \: ,
\end{equation}
in equation (\ref{eq:mu_i}). 

\begin{table}[bh]
\tbl{Quantum numbers of elementary particles and nuclei considered in the EoS models.}
{\begin{tabular}{@{}ccccccccccc@{}} \toprule
 quantum number & 
 $n$  & $\bar{n}$ & $p$ & $\bar{p}$ & $(A,Z)$ & $e^{-}$ & $e^{+}$ & $\mu^{-}$ & $\mu^{+}$ & $\gamma$ \\
\colrule
 $B_{i}$       & $+1$ & $-1$ & $+1$ & $-1$ & $A$ & 0 & 0 & 0 & 0 & 0 \\
 $Q_{i}$       & 0 &  0 & $+1$ & $-1$ & $Z$ & $-1$ & $+1$ & $-1$ & $+1$ & 0 \\
 $L_{i}^{(e)}$  & 0 &  0 & 0 & 0 & 0 & $+1$ & $-1$ & 0 & 0 & 0  \\
 $L_{i}^{(\mu)}$ & 0 &  0 & 0 & 0 & 0 & 0 & 0 & $+1$ & $-1$ & 0 \\
\botrule
\end{tabular}}
\label{tab:qn}
\end{table}


All nucleons, whether they are free or bound in nuclei, couple to the meson fields
with coupling factors $g_{ij}$ in equations (\ref{eq:source_v}) and (\ref{eq:source_s}).
Explicitly they are given by
\begin{eqnarray}
 g_{i\omega} & = & \left\{ \begin{array}{lll}
 B_{i} & \mbox{if} & i=n,p,\bar{n},\bar{p} \\
 A f(A)    & \mbox{if} & i = (A,Z)
 \end{array} \right.
 \\
 g_{i\sigma} & = & \left\{ \begin{array}{lll}
 |B_{i}| & \mbox{if} & i=n,p,\bar{n},\bar{p} \\
 A f(A)    & \mbox{if} & i = (A,Z) 
 \end{array} \right.
 \\
  g_{i\rho} & = & \left\{ \begin{array}{lll}
 B_{i} & \mbox{if} & i=n,\bar{n} \\
 -B_{i} & \mbox{if} & i=p,\bar{p} \\
 (A-2Z)f(A)    & \mbox{if} & i = (A,Z) 
 \end{array} \right.
 \\
  g_{i\delta} & = & 0
\end{eqnarray}
with a scaling function $f(A)$. It is defined as
\begin{equation}
 f(A) = \left\{ 
\begin{array}{lll}
 1 & \mbox{if} & A \leq 4 \\
 1-\left[ 1- \left( \frac{4}{A}\right)^{1/3} \right]^{3} 
 & \mbox{if} & A > 4
\end{array}
 \right. 
\end{equation}
in order to scale the coupling of heavy nuclei
according to their surface size. The factors $D_{i}$ in the density (\ref{eq:ntilde}),
which determine the density dependent coupling strengths $\Gamma_{j}$ in equation (\ref{eq:C_j}),
are identical to the baryon numbers $B_{i}$, hence $\tilde{n}$ is just the total baryon density
\begin{equation}
\label{eq:n_b}
 n_{b} = \sum_{i} B_{i} n_{i}^{(v)}
\end{equation}
of the system. The explicit form of the density dependent couplings in the DD2 parametrization
is described in Ref.\cite{Typel:2009sy}.

All particles are assumed to be pointlike, i.e., the available
volume fractions $\Phi_{i}$ are all constant equal to one and hence there are no rearrangement
contribution $V_{i}^{(r)}$ and $\omega_{\Phi}^{(r)}$ to the vector potentials (\ref{eq:V}) 
and the grand canonical potential density (\ref{eq:omega_b}), respectively.
Instead of the excluded volume mechanism, density and temperature dependent
mass shifts $\Delta m_{i}$ are introduced
in the scalar potential (\ref{eq:S}) in order to generate the dissolution of nuclei.
In contrast, explicit energy shifts $\Delta E_{i}$ are not introduced in the vector potential
(\ref{eq:V}).
The total mass shift includes two contributions
\begin{equation}
\label{eq:dm}
 \Delta m_{i} = \Delta m_{i}^{(\rm Coul)} + \Delta m_{i}^{(\rm strong)}
\end{equation}
in the gRDF model.
The Coulomb contribution is due to the screening of the 
Coulomb field by the electrons and muons.
It is given by
\begin{equation}
\label{eq:dm_c}
 \Delta m_{i}^{(\rm Coul)}(\{n_{i}^{(v)}\}) = - \frac{3}{5} \frac{Q_{i}^{2}\alpha}{R_{i}}
 \left( \frac{3}{2}x_{i} - \frac{1}{2} x_{i}^{3} \right)
\end{equation}
in Wigner-Seitz (WS) approximation 
for a nucleus $i=(A,Z)$ with radius $R_{i}$ that is estimated as
\begin{equation}
 R_{i} = R_{0} A_{i}^{1/3}
\end{equation}
using the radius parameter $R_{0} = 1.25$~fm.
The quantity $\alpha=e^{2}/(\hbar c)$ in (\ref{eq:dm_c}) denotes 
the fine structure constant. The ratio
\begin{equation}
 x_{i} = \frac{R_{i}}{R_{i,q}} 
\end{equation}
in equation (\ref{eq:dm_c})
contains the WS cell radius
\begin{equation}
 R_{i,q} = \left( - \frac{3Q_{i}}{4\pi 
  n_{q}^{(l)}} \right)^{1/3}
\end{equation}
with the total leptonic charge density
\begin{equation}
\label{eq:n_q_l}
 n_{q}^{(l)} = \sum_{i=e^{-},e^{+},\mu^{-},\mu^{+}} Q_{i} n_{i}
 = - n_{e^{-}} + n_{e^{+}} - n_{\mu^{-}} + n_{\mu^{+}} 
\end{equation}
that is negative.
Because the system is charge neutral, 
the negative of $n_{q}^{(l)}$ is identical to the
total hadronic charge density
\begin{equation}
\label{eq:n_q_h}
 n_{q}^{(h)} = \sum_{i=n,\bar{n},p,\bar{p},(A,Z)} Q_{i} n_{i}^{(v)} = Y_{q} n_{b}
\end{equation}
with the charge fraction $Y_{q}$.
The strong shift in equation (\ref{eq:dm}) is given by
\begin{equation}
 \Delta m_{i}^{(\rm strong)}(T,\{n_{i}^{(v)}\}) =  f_{i}(n_{i}^{(\rm eff)},n_{i}^{(\rm diss)}) 
 B_{i}^{(0)}
\end{equation}
with the binding energy $B_{i}^{(0)}$ of the nucleus in the vacuum  and a shift function
$f_{i}$, which depends on the effective density
\begin{equation}
 n_{i}^{(\rm eff)} 
  =\frac{2}{A_{i}} \left[ Z_{i} Y_{q} 
 + (N_{i}) (1-Y_{q})\right] n_{b} 
\end{equation}
and the dissolution density $n_{i}^{(\rm diss)}$.
The functional form of the shift function is chosen as
\begin{equation}
\label{eq:f_i}
 f_{i} = \left\{ \begin{array}{lll}
 x & \mbox{if} & x \leq 1 \\
 x+\frac{(x-1)^{3}(y-1)}{3(y-x)} & \mbox{if} & x > 1 \: \mbox{and} \: x < y
 \end{array} \right.
\end{equation}
with the two parameters
\begin{equation}
 x = \frac{n_{i}^{(\rm eff)}}{n_{i}^{(\rm diss)}} 
\end{equation}
and
\begin{equation}
 y = \frac{n_{\rm sat}}{n_{i}^{(\rm diss)}} \: .
\end{equation}
For light nuclei, the dissolution density is 
\begin{equation}
 n_{i}^{(\rm diss)} = \frac{B_{i}^{(0)}}{\delta B_{i}(T)}
\end{equation}
where the quantity $\delta B_{i}(T)$ is defined in Ref.\cite{Typel:2009sy}.
For nucleon-nucleon scattering states, the shift function of the deuteron is used.
For heavy nuclei, the parametrisation
\begin{equation}
 x = \frac{n_{i}^{(\rm eff)}}{n_{\rm sat}} y
\end{equation}
with
\begin{equation}
 y = 3+\frac{28}{A}
\end{equation}
is used such that
\begin{equation}
  n_{i}^{(\rm diss)} = \frac{n_{\rm sat}}{y} \: .
\end{equation}
For $A=4$, the approximate dissolution density of $\alpha$-particles at zero
temperature is reproduced.
With increasing mass number $A$, the dissolution density $n_{i}^{(\rm diss)}$ 
approaches one third of the saturation density. The shift function (\ref{eq:f_i})
replaces the corresponding function in equation (72) of Ref.\cite{Typel:2009sy}. 
In the present form, the binding energy
of a nucleus decreases linearly with the density until the binding threshold is reached.
At higher densities, $f_{i}$ grows more rapidly since the state becomes a resonance.
The strong increase also avoids the reappearance of nuclei, in particular light clusters,
at very high densities.

The degeneracy factors $g_{i}$ of the elementary particles and light nuclei
($d$, $t$, $h$, $\alpha$) are given by
\begin{equation}
 g_{i} = 2J_{i}+1
\end{equation}
with the spin $J_{i}$ of the particle (in the ground state). The contribution of excited states
is considered for heavy nuclei with $A>4$ by introducing temperature dependent
degeneracies
\begin{eqnarray}
\label{eq:g_i_T}
g_{i}(T) =  \left[ 2J_{i}+1 
 + \int_{0}^{E_{\rm max}} d\varepsilon \: 
 \varrho_{i}(\varepsilon) \exp \left( -\frac{\varepsilon+\Delta_{i}}{T} \right)\right]
 \gamma(T)
\end{eqnarray}
with two factors. The first term contains the contribution of the ground state and of
excited states with a density of states $\varrho_{i}(\varepsilon)$. The
second factor $\gamma(T)$ is introduced to model the dissolution of heavy nuclei 
at high temperatures. Experimental values for the ground state spin $J_{i}$ 
of nuclei are used as far as known. Otherwise $J_{i}=0$, $J_{i} = 1/2$ or $J_{i} = 1$ are assumed
for even-even, odd-even/even-odd or odd-odd nuclei, respectively.
The density of states is given by a modified Fermi gas formula\cite{GrFe:1985}
\begin{eqnarray}
\label{eq:rho_i}
\varrho_{i}(\varepsilon)=\frac{\sqrt{\pi}}{24}
 \frac{a_{i}}{\sqrt{a_{i}^{(n)} a_{i}^{(p)}}}
 \frac{\exp{\left(\beta_{i}\varepsilon
 +\frac{a_{i}}{\beta_{i}}\right)}}{\left(\beta_{i}\varepsilon^{3}\right)^{1/2}}
 \frac{1-\exp{\left(-\frac{a_{i}}{\beta_{i}}\right)}}{\left[1-\frac{1}{2}\beta_{i}
 \varepsilon\exp\left(-\frac{a_{i}}{\beta_{i}}\right)\right]^{1/2}}
\end{eqnarray}
with the level density parameters 
\begin{equation}
 a_{i}^{(j)}=\frac{\pi^{2}}{3}g_{i}^{(j)}
\end{equation}
for neutrons and protons ($j=n,p$)
in a nucleus $i$ and their sum $a_{i} = a_{i}^{(n)}+a_{i}^{(p)}$. The factor 
\begin{equation}
 g_{i}^{(j)}=\frac{m_{j} k_{i}^{(j)}}{2\pi^{2}}
\end{equation}
depends on the Fermi momenta
\begin{equation}
 k_{i}^{(j)} = \left[ \frac{3\pi^{2}N_{i}^{(j)}}{V_{i}} \right]^{1/3}
\end{equation}
with the number of nucleons ($N_{i}^{(n)} = N$, $N_{i}^{(p)} = Z$) in the nucleus
$i=(A,Z)$ and its effective volume
\begin{equation}
 V_{i} = \frac{4\pi}{3} R_{i}^{3}
\end{equation}
with radius $R_{i} = 1.4 \cdot A^{1/3}$~fm.
For given $a_{i}$ and $\epsilon$, the ratio $a_{i}/\beta_{i}$ and thus $\beta_{i}$ are determined by solving
the equation
\begin{equation}
 \left( \frac{a_{i}}{\beta_{i}} \right)^{2} =
 a_{i} \varepsilon\left[1-\exp{\left(-\frac{a_i}{\beta_i}\right)}\right] \: .
\end{equation}
Then all quantities in equation (\ref{eq:rho_i}) are known and the density of states can be calculated.
In contrast to usual Fermi gas models, $\varrho_{i}(\epsilon)$ does not diverge for $\varepsilon \to 0$
but assumes a finite value. The maximum energy $E_{\rm max}$ in (\ref{eq:rho_i}) is given by the minimum
of the neutron and proton separation energies of the nucleus and $\Delta_{i}=\delta_{i}/A^{1/3}$~MeV 
is a simple pairing correction with $\delta_{i} = 0,1,2$
for even-even, odd-even/even-odd and odd-odd nuclei, respectively.
The dissolution factor in equation (\ref{eq:g_i_T}) is written as
\begin{equation}
 \gamma(T) = \left\{ \begin{array}{lll}
     \exp \left[ - \left( \frac{T}{T_{\rm fl}-T}\right)^{2}\right]
 & \mbox{if} & T < T_{\rm fl} \\
 0 & \mbox{if} & T \geq T_{\rm fl}
 \end{array} \right.
\end{equation}
with the flashing temperature $T_{\rm fl} = 11.26430$~MeV of the DD2-RMF model. 
It is defined
by the condition that the pressure  $p(T,n_{b})$ attains a minimum of $p=0$ 
as a function of the baryon density $n_{b}$ for constant 
temperature in symmetric nuclear matter.
$T_{\rm fl}$ is slightly lower than
the critical temperature $T_{\rm cr}=13.72384$~MeV 
of the liquid-gas phase transition at which
$\left. \partial p/\partial n_{b} \right|_{T_{\rm cr}}=0$. 
For $T< T_{\rm cr}$ uniform nuclear matter
is globally unstable
and for $T < T_{\rm fl}$ locally unstable to density fluctuations.

Because the system is presumed to be charge neutral, the leptonic charge density
(\ref{eq:n_q_l}) has to compensate the hadronic charge density (\ref{eq:n_q_h}) such that
$n_{q}^{(l)}+n_{q}^{(h)}=0$. This condition reduces the dimension of the thermodynamic space
of variables and determines the leptonic chemical potential $\mu_{l}$.
In order to find the thermodynamic properties of the ensemble for given baryon density $n_{b}$ and
hadronic charge fraction $Y_{q}$, the remaining independent chemical potentials $\mu_{b}$ and $\mu_{q}$ are
varied such that $n_{b}$ and $Y_{q}$ are reproduced when the coupled set of equations for the
densities and potentials are solved self-consistently. Predictions of the gRDF model
for a number of thermodynamic quantities and for the chemical composition of matter are available
in tabular form according to the CompOSE format \cite{compose}. 
The table covers temperatures from $0.1$~MeV
to $100$~MeV, baryon densities from $10^{-10}$~fm$^{-3}$ to $1$~fm$^{-3}$, and hadronic charge fractions
from $0.01$ to $0.60$.

\subsection{HS model}
\label{sec:HS}

The statistical model of Hempel and Schaffner-Bielich\cite{Hempel:2009mc}, denoted HS model in
the following, is formulated using a canonical ensemble
of nucleons, nuclei, electrons and photons by specifying the free energy density
\begin{equation}
\label{eq:f}
 f(T,\{n_{i}\}) = \omega(T,\{\mu_{i}\}) + \sum_{i} \mu_{i}n_{i} \: .
\end{equation}
It can be obtained from the grand canonical potential density (\ref{eq:omega}) by applying
a Legendre transformation to exchange the chemical potentials $\mu_{i}$ with the particle
number densities $n_{i}$ as natural variables. There are different variants of the HS 
model\cite{Hempel:2009mc,Hempel:2011mk,Steiner:2012rk,Fischer:2013eka},
depending on the choice of the parametrization of the nuclear interaction 
and the table of nuclei.
Here we consider the version of Ref.\cite{Fischer:2013eka} with the DD2 parametrization.

The set of particle degrees of freedom is
similar to that of the gRDF model, however, muons and two-nucleon 
scattering correlations are not
taken into account in the HS model. An extensive table of nuclei is considered with 
masses taken from the 2003 Atomic Mass Evaluation\cite{Audi:2003} if available or otherwise 
from the microscopic-macroscopic FRDM model\cite{Moller:1993ed}. 
Effects of the nuclear interaction are included
in a RMF model with density-dependent couplings in the DD2 parametrization\cite{Typel:2009sy}
identical to the gRDF approach of the previous section. 
However, in the HS model, nucleons inside
nuclei do not couple to the mesons, i.e., the coupling factors 
in equations (\ref{eq:source_v}) and (\ref{eq:source_s}) are defined as
\begin{eqnarray}
 g_{i\omega} & = & \left\{ \begin{array}{lll}
 B_{i} & \mbox{if} & i=n,p,\bar{n},\bar{p} \\
 0    & \mbox{if} & i = (A,Z)
 \end{array} \right.
 \\
 g_{i\sigma} & = & \left\{ \begin{array}{lll}
 |B_{i}| & \mbox{if} & i=n,p,\bar{n},\bar{p} \\
 0    & \mbox{if} & i = (A,Z) 
 \end{array} \right.
 \\
  g_{i\rho} & = & \left\{ \begin{array}{lll}
 B_{i} & \mbox{if} & i=n,\bar{n} \\
 -B_{i} & \mbox{if} & i=p,\bar{p} \\
 0    & \mbox{if} & i = (A,Z) 
 \end{array} \right.
 \\
  g_{i\delta} & = & 0 \: .
\end{eqnarray}
Correspondingly, only nucleons are considered in the sum (\ref{eq:ntilde}) with $D_{i} = B_{i}$
that appear as argument $\tilde{n}$ in the meson-nucleon couplings $\Gamma_{j}$.

For the degeneracy factors $g_{i}$ of the nuclei, the form
\begin{eqnarray}
\label{eq:g_i_T_HS}
g_{i}(T) =   2J_{i}+1 
 + \int_{0}^{E_{\rm max}} d\varepsilon \: 
 \varrho_{i}(\varepsilon) \exp \left( -\frac{\varepsilon}{T} \right)
\end{eqnarray}
is employed with the empirical density of states
\begin{equation}
 \varrho_{i}(\varepsilon) = \frac{c_{1}}{A^{5/3}} \exp \left( \sqrt{2a(A)\varepsilon} \right)
\end{equation}
containing the mass dependent level density parameter 
\begin{equation}
 a(A) = \frac{A}{8} \left( 1 - \frac{c_{2}}{A^{1/3}} \right) \: \mbox{MeV}^{-1}
\end{equation}
and constants $c_{1} = 0.2$~MeV$^{-1}$ and $c_{2} = 0.8$.
The maximum energy $E_{\rm max}$ for the integration in (\ref{eq:g_i_T_HS}) 
is given by the binding energy of the
specific nucleus such that only bound excited states are taken into account.
The ground state spins in (\ref{eq:g_i_T_HS}) are taken as $J_{i}=0$ and $J_{i} = 1$ 
for even A and odd A nuclei, respectively.

The HS model uses the concept of finite sizes for nucleons and nuclei with two realizations
of the excluded-volume mechanism in order to incorporate the dissolution of nuclei with increasing
density of the system. All other particles (electrons, photons) are treated as pointlike.
For nuclei $i=(A,Z)$ the available volume fractions
are given by the standard form (\ref{eq:phi_i}) with volume
parameters $v_{ij} = B_{j}/n_{\rm sat}$ ($j=$ nucleon or nucleus) 
proportional to the baryon number $B_{j}$. This 
leads to identical forms
\begin{equation}
 \Phi_{(A,Z)} = \kappa = \left\{ \begin{array}{lll}
 1 - \frac{n_{b}}{n_{\rm sat}} & \mbox{if} & n_{b} \leq n_{\rm sat} \\
 0 & \mbox{if} & n_{b} > n_{\rm sat}
\end{array} \right.
\end{equation}
with the baryon density (\ref{eq:n_b}) independent of the nucleus $(A,Z)$.
The available volume fraction $\Phi_{(A,Z)}$ vanishes for $n_{b} \geq n_{\rm sat}$
and hence the occurence of nuclei at densities above saturation is prohibited.
For nucleons $i=n,p,\bar{n},\bar{p}$ the filling factor
\begin{equation}
\label{eq:xi}
 \xi = 1 - \sum_{(A,Z)} A \frac{n_{(A,Z)}}{n_{\rm sat}}
\end{equation}
is introduced in the HS model where only nuclei contribute in the sum. 
Hence, nucleons do not block the volume
that is available for themselves. They only feel the occurrence of nuclei. However, the factor
$\xi$ cannot be considered as an available volume fraction in the sense of section
\ref{sec:exvol} and thus we set $\Phi_{i}=1$ for nucleons in the following.

As in the gRDF model, the interaction of the nuclei with the background electrons 
is taken into account in the HS model using the Wigner-Seitz approximation.
This can be formulated as a mass shift, which is given by equation (\ref{eq:dm_c}).
However, in the HS model, the mass shift $\Delta m_{i}$ is replaced by an
explicit energy shift for nuclei
\begin{equation}
\label{eq:dE}
 \Delta E_{i}(\{n_{i}^{(v)}\}) = - \frac{3}{5} \frac{Q_{i}^{2}\alpha}{R_{i}}
 \left( \frac{3}{2}x_{i} - \frac{1}{2} x_{i}^{3} \right)
\end{equation}
of the same form as in equation (\ref{eq:dm_c}). The radius of a nucleus is calculated
from
\begin{equation}
 R_{(A,Z)} = \left( \frac{3 A}{4\pi n_{\rm sat}} \right)^{1/3} \: .
\end{equation}
Since the Coulomb energy shift (\ref{eq:dE}) depends only on the electron and positron
densities, there are only energy rearrangement contributions for leptons.
Mass shifts of nuclei are not considered in the HS model.

Using the formalism of section \ref{sec:exvol}, the free energy density 
(\ref{eq:f}) of the HS model can be deduced.
Since the available volume fractions $\Phi_{i}$ are either constant one or assume the standard
form (\ref{eq:phi_i}), the simplified form (\ref{eq:omega_simple}) of the grand canonical
potential density can be used to obtain $f(T,\{n_{i}\})$.
It is convenient to examine the contributions of different particles to the free energy density
separately, i.e., we write
\begin{equation}
\label{eq:f_HS}
  f(T,\{n_{i}\}) = \sum_{(A,Z)}f_{(A,Z)} + f_{\rm nuc} + f_{e} + f_{\gamma}
\end{equation}
with the contribution of nuclei, nucleons (including mesons), electrons, and photons. 
Since nuclei are treated as nonrelativistic particles with Maxwell-Boltzmann statistics,
we can use (\ref{eq:n_i_MB}) and solve it
for the chemical potential
\begin{equation}
 \mu_{i} = T \ln \left[ \frac{n_{i} \left(\lambda_{i}^{\ast}\right)^{3}}{g_{i}\Phi_{i}}
 \right] + m_{i}^{\ast} + V_{i} \: .
\end{equation} 
We note that $V_{i} = \Delta E_{i}$ and $m_{i}^{\ast} = m_{i}$ for nuclei.
The first term in equation (\ref{eq:f_HS}) is given by
\begin{eqnarray}
\label{eq:f_AZ}
 \sum_{(A,Z)}f_{(A,Z)} & = & \sum_{(A,Z)} \left( \omega_{(A,Z)}^{(0)} 
 + \mu_{(A,Z)} n_{(A,Z)}\right) 
 \\ \nonumber & = & 
 \sum_{(A,Z)} \left\{  T \ln \left[ \frac{n_{(A,Z)}
 \lambda_{(A,Z)}^{3}}{g_{(A,Z)}\Phi_{(A,Z)}}
 \right] - T + m_{(A,Z)} + V_{i} \right\}  n_{(A,Z)} 
 \\ \nonumber & = & 
 \sum_{(A,Z)} f_{(A,Z)}^{(0)}(T,n_{(A,Z)})
 - T \sum_{(A,Z)} n_{(A,Z)} \ln \kappa 
 + f_{\rm Coul}
\end{eqnarray}
with the free energy density
\begin{equation}
 f_{i}^{(0)}(T,n_{i}) = \left[ T \ln \left( \frac{n_{i}
 \lambda_{i}^{3}}{g_{i}}
 \right) - T + m_{i} \right]  n_{i} 
\end{equation}
for an ideal gas of particles $i$ with rest mass $m_{i}$ and thermal wavelength 
$\lambda_{i} = \sqrt{2\pi/(m_{i}T)}$. 
The explicit excluded-volume term in (\ref{eq:f_AZ}) 
contains the available volume fraction $\kappa = \Phi_{(A,Z)}$.
The Coulomb term
\begin{equation}
 f_{\rm Coul}  =   
 \sum_{(A,Z)} n_{(A,Z)} \Delta E_{(A,Z)}^{\rm (Coul)}
\end{equation}
arises due to the energy shifts (\ref{eq:dE}). 

The nucleonic contribution to (\ref{eq:f_HS})
\begin{equation}
 f_{\rm nuc}(T,n_{n},n_{p},n_{\bar{n}},n_{\bar{p}}) 
 = \sum_{i=n,p,\bar{n},\bar{p}} \left( \omega_{i}^{(0)} + \mu_{i} n_{i} \right) 
 + \omega_{\rm meson} - \omega_{\rm meson}^{(r)}
\end{equation}
depends only on the temperature and the densities of the nucleons. This form is not used
in the HS model directly but
the rescaled contribution
\begin{equation}
\label{eq:xif}
 \xi f_{\rm nuc}(T,n_{n}^{\prime},n_{p}^{\prime},n_{\bar{n}}^{\prime},n_{\bar{p}}^{\prime}) 
\end{equation}
with the scaled nucleon densities
\begin{equation}
 n_{i}^{\prime}  = \frac{n_{i}}{\xi}
\end{equation}
by introducing the filling factor (\ref{eq:xi}). If nuclei exist in the system, the available
volume for the nucleons is reduced, e.i., $\xi < 1$. Correspondingly, the effective density
in the available volume has to be increased as compared to the densities in the total volume
and the obtained results for the free energy density
needs to be adjusted again to the total volume leading to the prefactor $\xi$
in equation (\ref{eq:xif}).

The screening of the Coulomb potential due to the electrons affects not only the energies of nuclei
but also the thermodynamics of the electrons and positrons themselves. 
The scalar potential $S_{i}$ vanishes
but the vector potential is given by a rearrangment term
\begin{eqnarray}
V_{i} & = & W_{i}^{(r)} = \sum_{(A,Z)} n_{(A,Z)} \frac{\partial \Delta E_{(A,Z)}}{\partial n_{i}}
 \\ \nonumber & = & 
 \mp \sum_{(A,Z)} n_{(A,Z)} 
 \frac{3}{10} \frac{Z^{2}\alpha}{R_{(A,Z)} n_{e}}
 \left[ x_{(A,Z)}-  x_{(A,Z)}^{3} \right]
\end{eqnarray}
with the net electron number density $n_{e} = n_{e^{-}}-n_{e^{+}}$.
The negative (positive) sign applies to electrons (positrons).
These particles get effective chemical potentials $\mu_{i}^{\ast} = \mu_{i} -V_{i}$
with the correct sign change of $\mu_{i}$ and $V_{i}$ for the particle to antiparticle
conversion. The electronic contribution to the total free energy
density assumes the form
\begin{equation}
 f_{e} = \sum_{i=e^{-},e^{+}} \left( \omega_{i}^{(0)}  + \mu_{i} n_{i} \right) 
 - \omega_{\Delta}^{(r)}
\end{equation}
with the explicit rearrangement term
\begin{equation}
 \omega_{\Delta}^{(r)} = \sum_{i=e^{-},e^{+}} n_{i} W_{i}^{(r)}
 = - \sum_{(A,Z)} n_{(A,Z)} 
 \frac{3}{10} \frac{Z^{2}\alpha}{R_{(A,Z)}}
 \left[ x_{(A,Z)}-  x_{(A,Z)}^{3} \right] \: .
\end{equation}

\section{Results for neutron star matter}
\label{sec:res}

\begin{figure}[ht]
  \centerline{
    \subfigure[]{\includegraphics[width=0.48\textwidth]{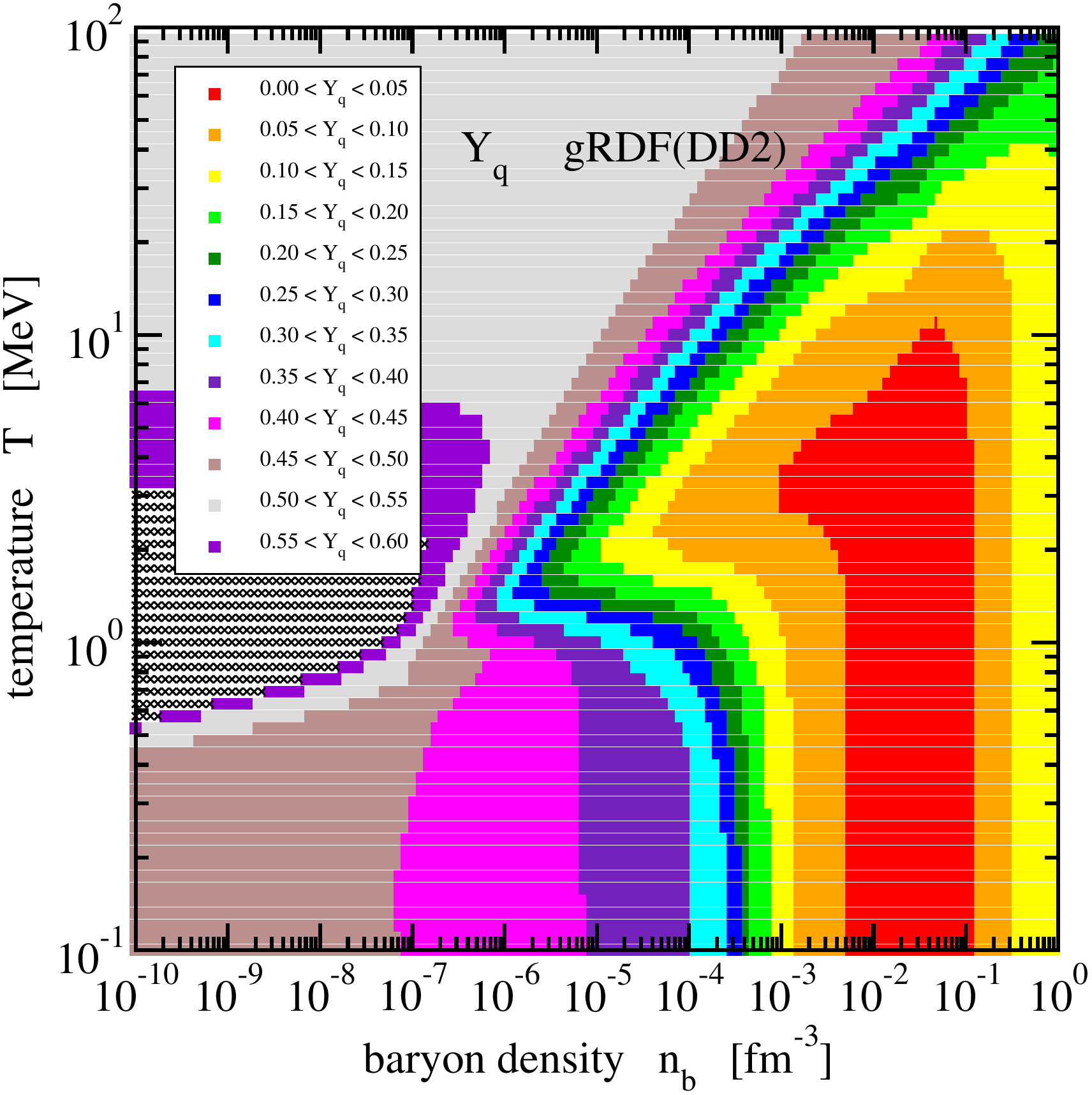}}
    \hspace*{4pt}
    \subfigure[]{\includegraphics[width=0.48\textwidth]{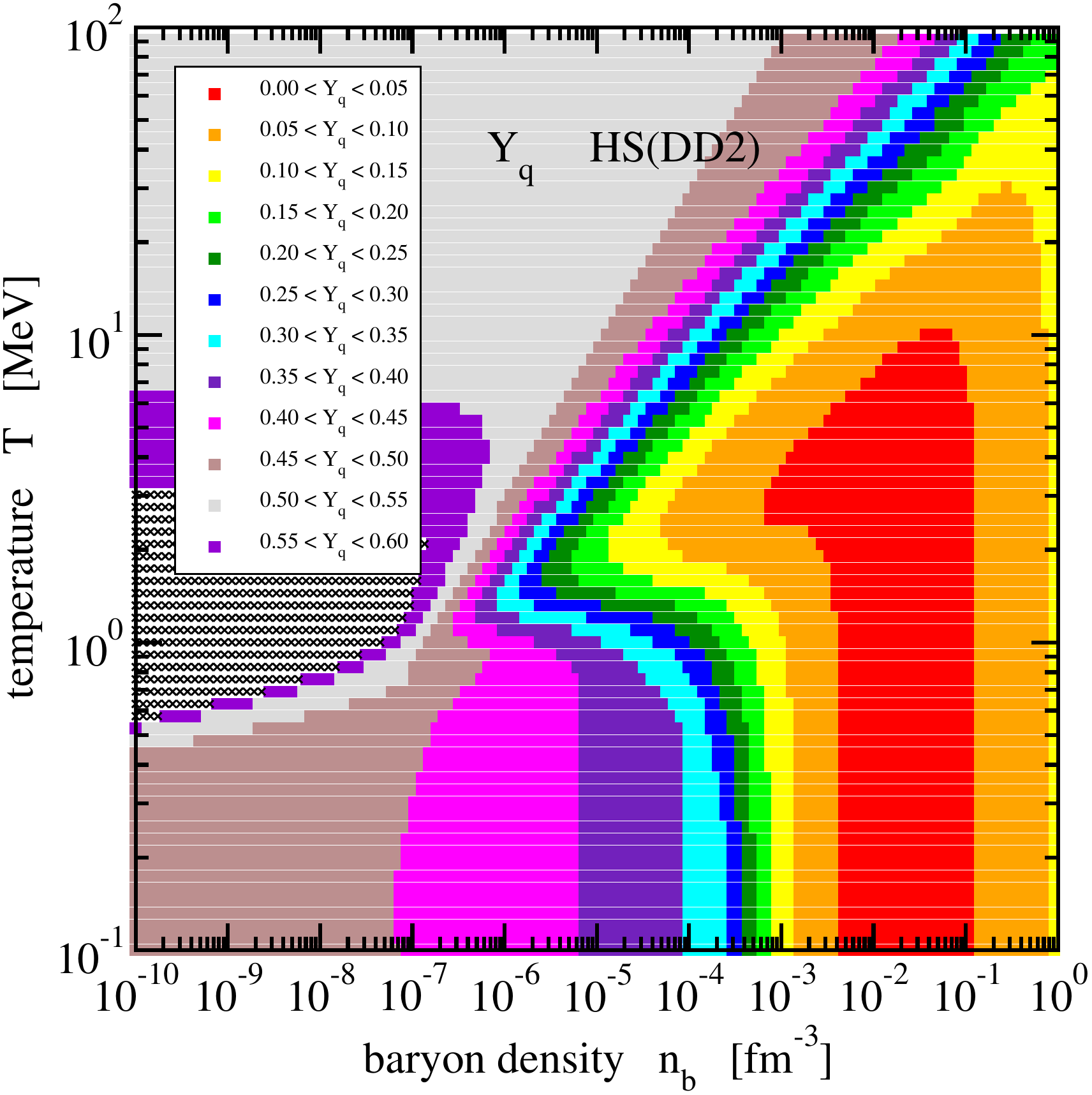}}
  }
  \caption{(Color online) Hadronic charge fraction $Y_{q}$ of neutron star matter
    for the gRDF model (a) and the HS model (b). The area of datapoints outside the EoS table 
    ($Y_{q}$ larger than $0.6$) is indicated by crosses.}
  \label{fig:01}
\end{figure}

Properties of stellar matter are available for the gRDF and HS model for a large range
of densities, temperatures and isospin asymmetries. In the present study they were extracted
from EoS tables in the CompOSE format using the {\sc FORTRAN} code {\tt compose.f90}.
In order to compare the two models, only 
a section through the full space of these variables is considered in the following. We restrict
ourselves to the conditions of fully catalyzed neutron star matter, i.e., charge-neutral matter in 
$\beta$ equilibrium without neutrinos. Then, the hadronic charge fraction $Y_{q}$ is determined and
the baryon density $n_{b}$ and the temperature $T$ remain as independent variables. 
In figure \ref{fig:01} the variation of $Y_{q}$ as a function of $n_{b}$ and $T$ is depicted
for the gRDF and the HS model. Regions of different charge fraction values are color-coded
in steps of 0.05. For baryon densities below approx.\ $10^{-7}$~fm${}^{-3}$ and temperatures between
$0.5$~MeV and $3$~MeV, $Y_{q}$ is above $0.6$ and thus outside the range of the EoS tables.
In general, there is a rather good agreement of the predictions between the two models.
For constant temperature, there is a decrease of the charge fraction with increasing baryon density
as long as $n_{b}$ is lower than approx.\ $4 \cdot 10^{-2}$~fm${}^{-3}$. For $T > 10$~MeV a further
decrease of $Y_{q}$ with $n_{b}$ is observed, whereas the charge fraction increases at lower $T$
for the highest baryon densities. Above the nuclear saturation density, a somewhat larger
difference in $Y_{q}$ between the models can be noticed 
because the gRDF model considers muons in contrast 
to the HS model. The lowest charge fractions with $Y_{q} < 0.05$ are found for $T < 10$~MeV 
in the baryon density range from roughly $5 \cdot 10^{-3}$~fm${}^{-3}$ to $10^{-1}$~fm${}^{-3}$.
At the lowest temperatures, the hadronic charge fraction evolves with $n_{b}$ as expected
for cold neutron stars indicating the gradual neutronization of matter with increasing depth.

\subsection{Chemical composition}

\begin{figure}[t]
  \centerline{
    \subfigure[]{\includegraphics[width=0.48\textwidth]{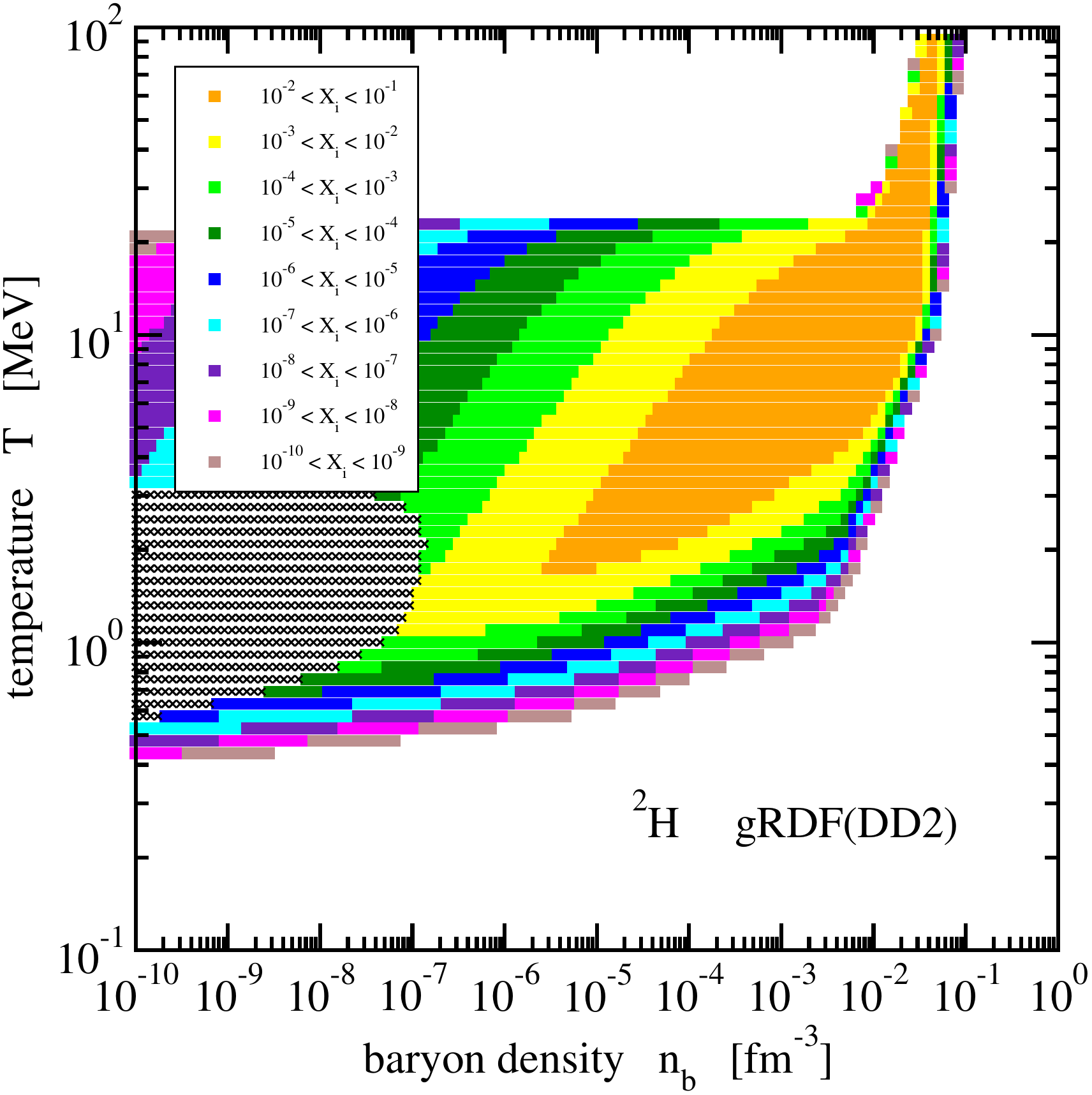}}
    \hspace*{4pt}
    \subfigure[]{\includegraphics[width=0.48\textwidth]{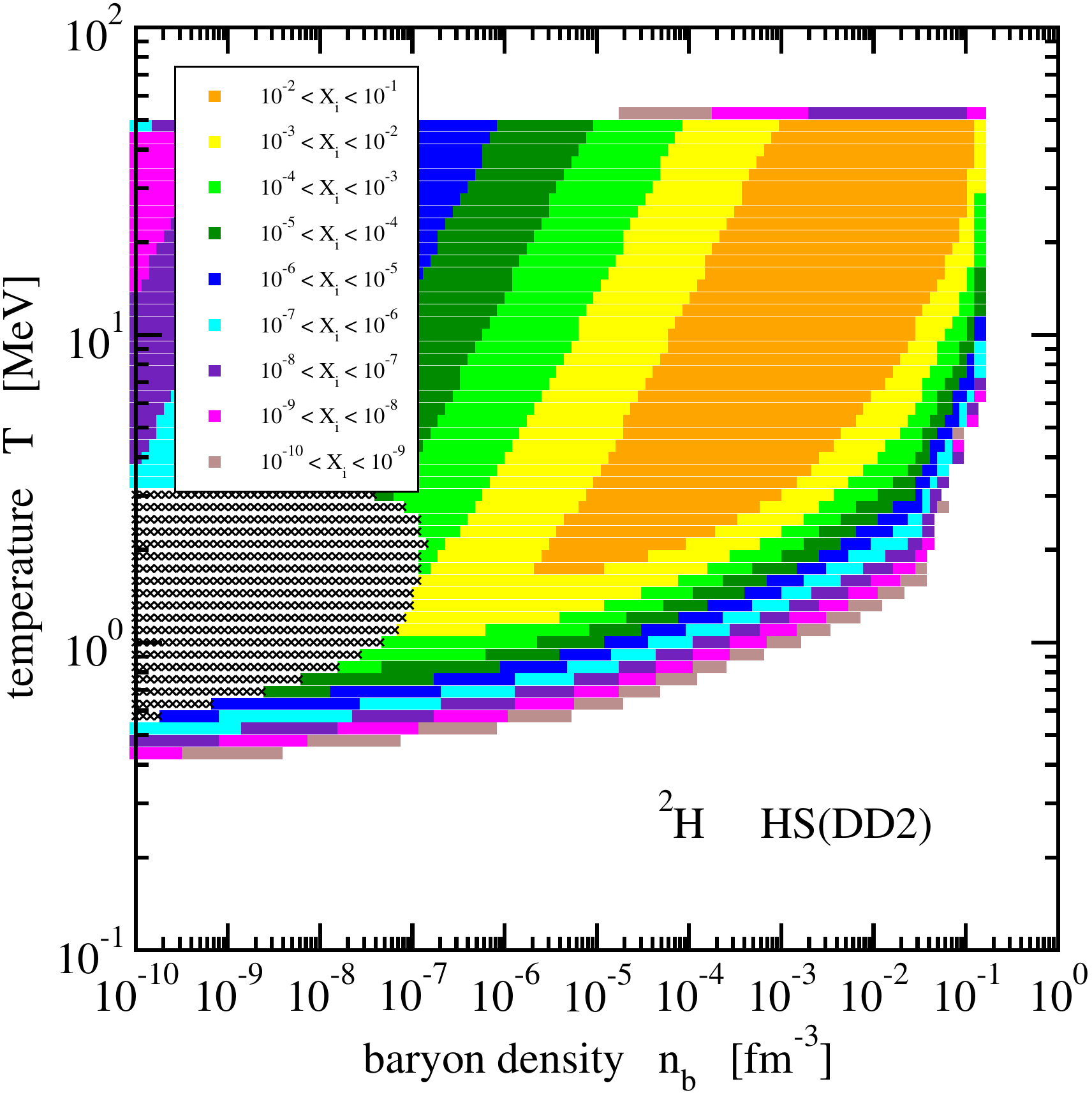}}
  }
  \centerline{
    \subfigure[]{\includegraphics[width=0.48\textwidth]{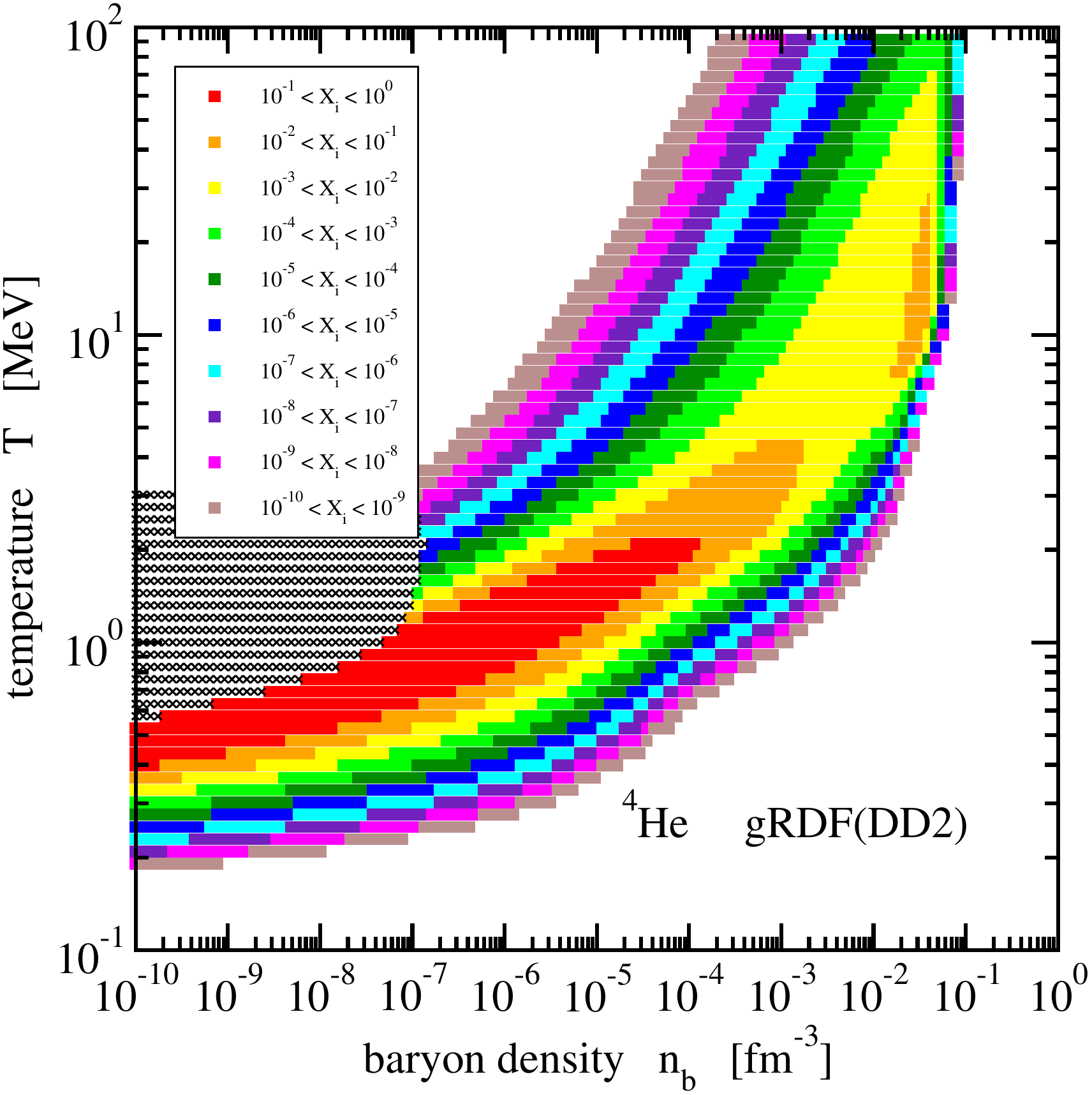}}
    \hspace*{4pt}
    \subfigure[]{\includegraphics[width=0.48\textwidth]{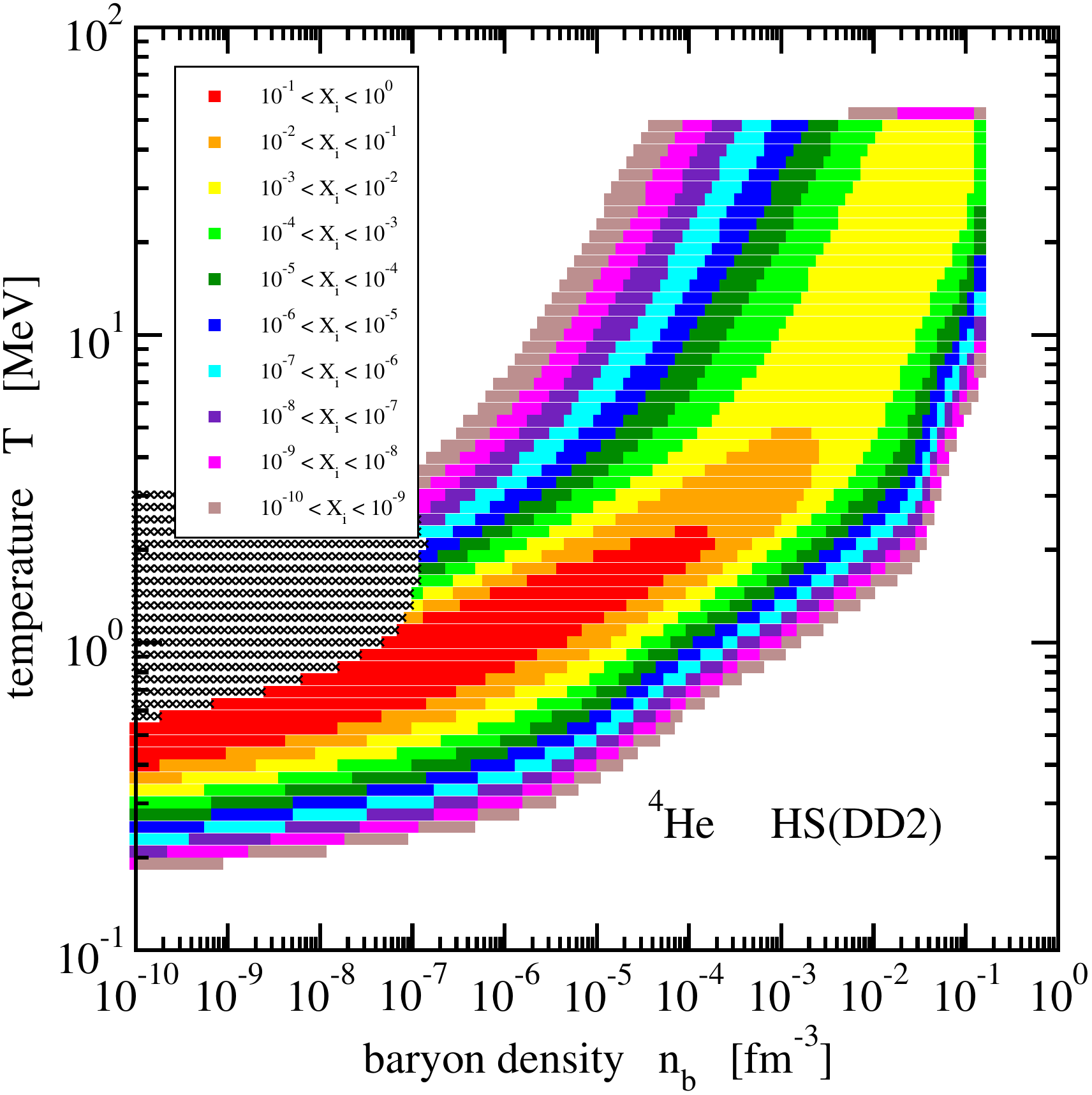}}
  }
  \caption{(Color online) Mass fraction $X_{d}$ of deuterons (${}^{2}$H)
    for the gRDF model (a) and the HS model (b) and 
    mass fraction $X_{\alpha}$ of $\alpha$ particles (${}^{4}$He)
    for the gRDF model (c) and the HS model (d) in neutron star matter.
    The area of datapoints outside the EoS table is indicated by crosses.}
  \label{fig:02}
\end{figure}

The change of the hadronic charge fraction is accompanied with a change of the chemical 
composition in neutron star matter. The gRDF and HS models predict to some extent differences 
for both light and heavy nuclei. We first consider the isospin zero light nuclei.
In figure \ref{fig:02}  the  mass fractions
\begin{equation}
 X_{i} = A_{i} \frac{n_{i}}{n_{b}}
\end{equation}
of deuterons and $\alpha$ particles, respectively,  
are depicted in the full range of baryon densities and temperatures. 
The contour lines of the color-coded regions denote a factor of ten change of the
fractions. For temperatures below
approx.\ $10$~MeV, an agreement of the two models is found to a large extent. 
In most cases, the $\alpha$-particle fraction exceeds that of the deuteron, except for
densities above $10^{-2}$~fm${}^{-3}$ and temperatures larger than approx.\ $4$~MeV.
The maximum fractions of deuterons and, in particular $\alpha$ particles, are located in a
rather narrow band of increasing density and temperature.
At temperatures $T > 20$~MeV, the largest differences between the models
are observed. In the gRDF model, the net deuteron fraction decreases quickly with increasing temperature, for 
baryon densities $n_{b} < 10^{-2}$~fm${}^{-3}$, because
the ground state contribution is compensated by the contribution of the corresponding
scattering channel resulting in a disappearance of deuterons at higher temperatures.
Contributions from two-nucleon scattering states are not considered in the HS model. Hence, the 
deuteron survives to much higher temperatures. For $\alpha$-particles, a compensating effect
due to scattering states is neglected in both models because the ground state is much stronger
bound as compared to the deuteron. In the HS model, the occurrence of clusters is suppressed
at temperatures above $50$~MeV by introducing an artificial cutoff. This is most apparent
for the $\alpha$-particle mass fractions in the bottom of figure \ref{fig:02}.

\begin{figure}[ht]
  \centerline{
    \subfigure[]{\includegraphics[width=0.48\textwidth]{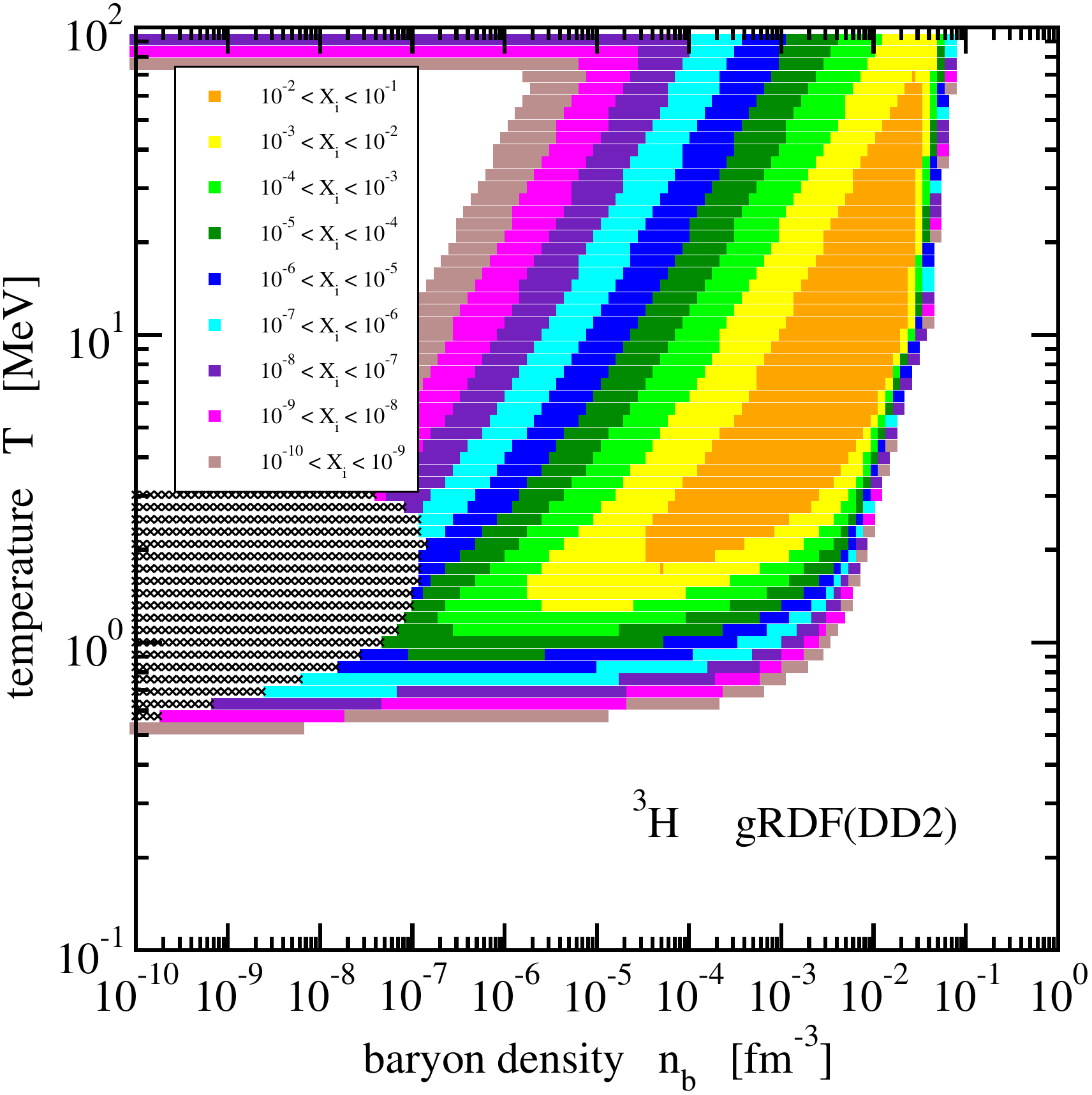}}
    \hspace*{4pt}
    \subfigure[]{\includegraphics[width=0.48\textwidth]{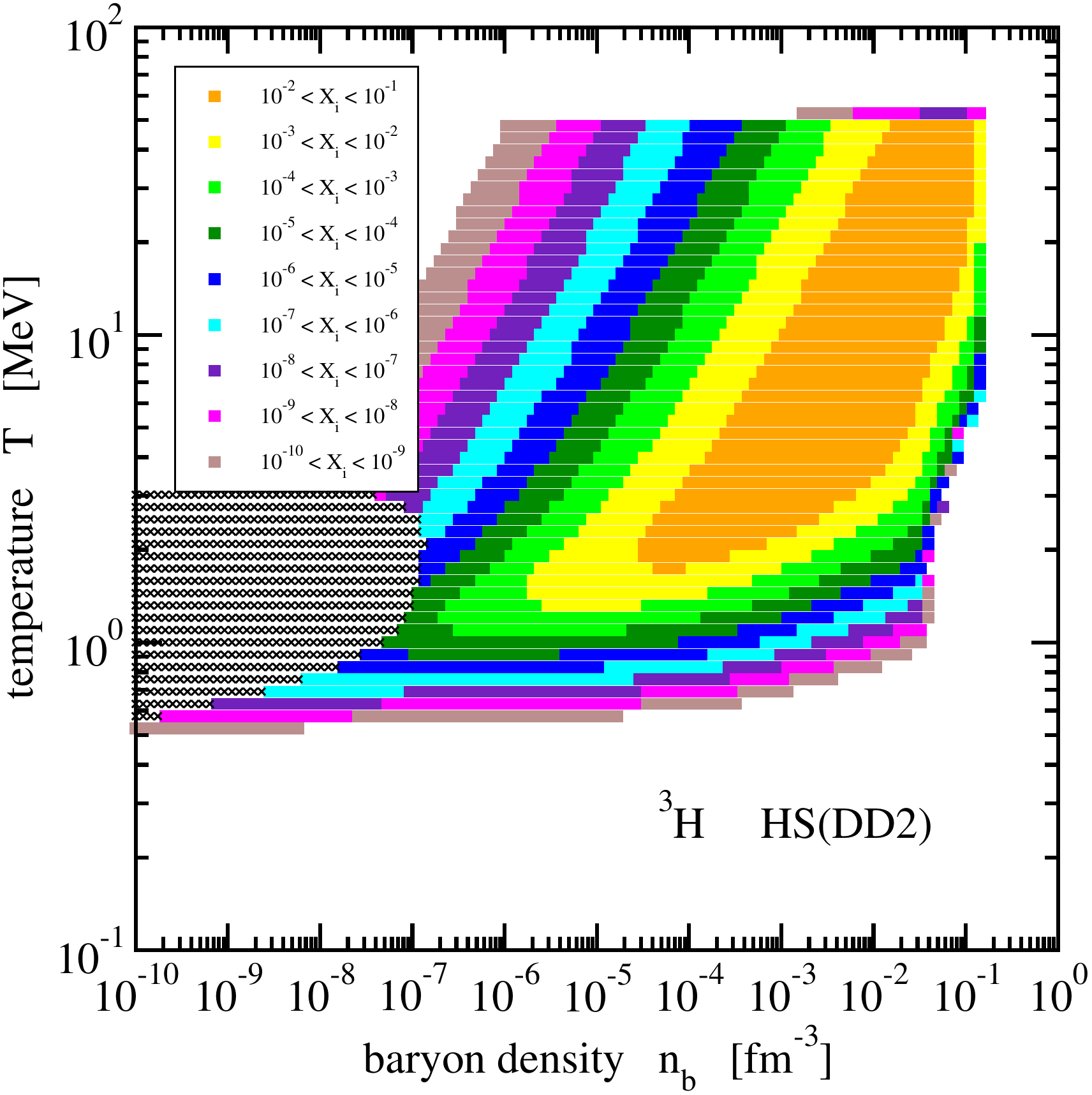}}
  }
  \centerline{
    \subfigure[]{\includegraphics[width=0.48\textwidth]{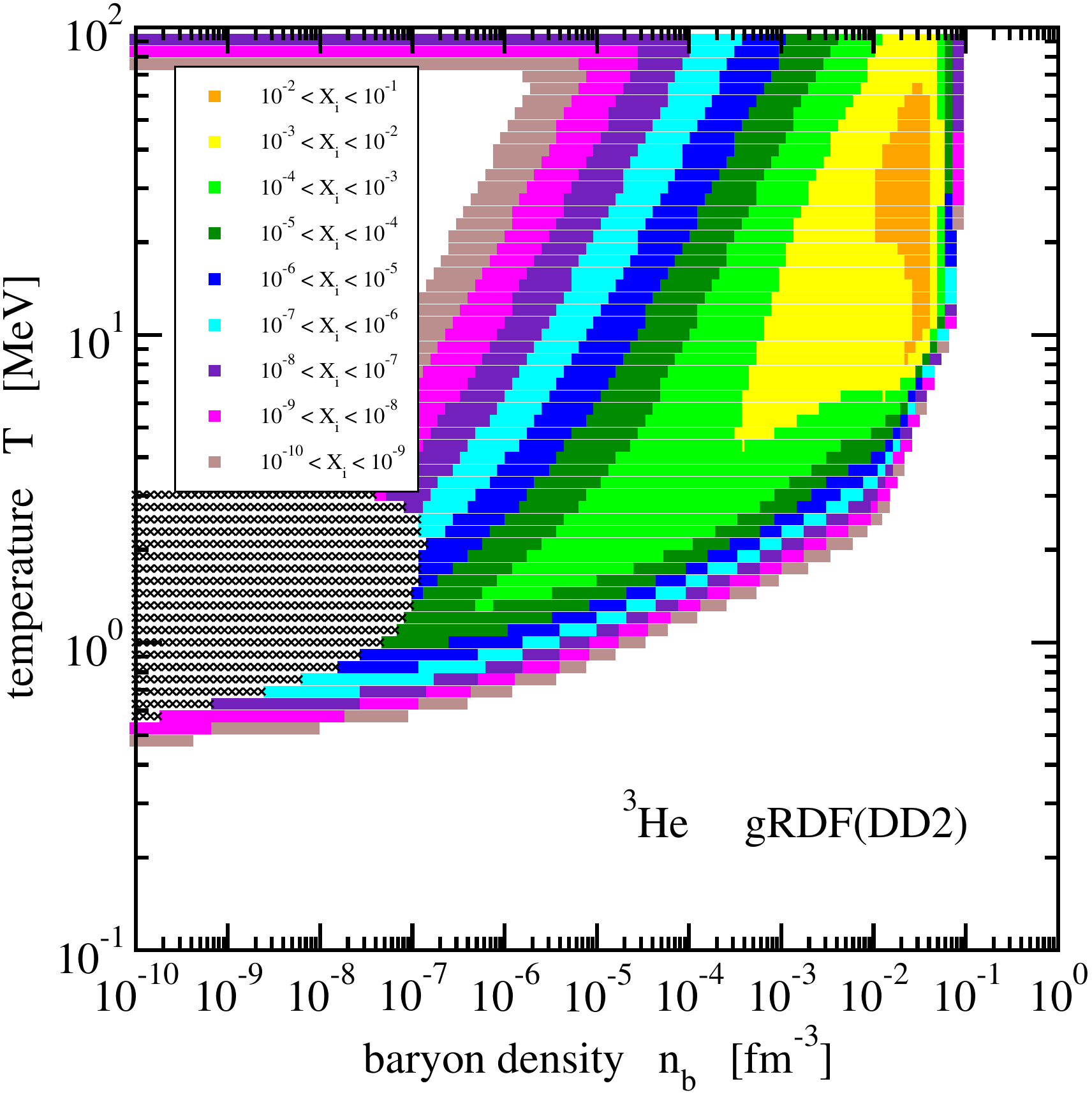}}
    \hspace*{4pt}
    \subfigure[]{\includegraphics[width=0.48\textwidth]{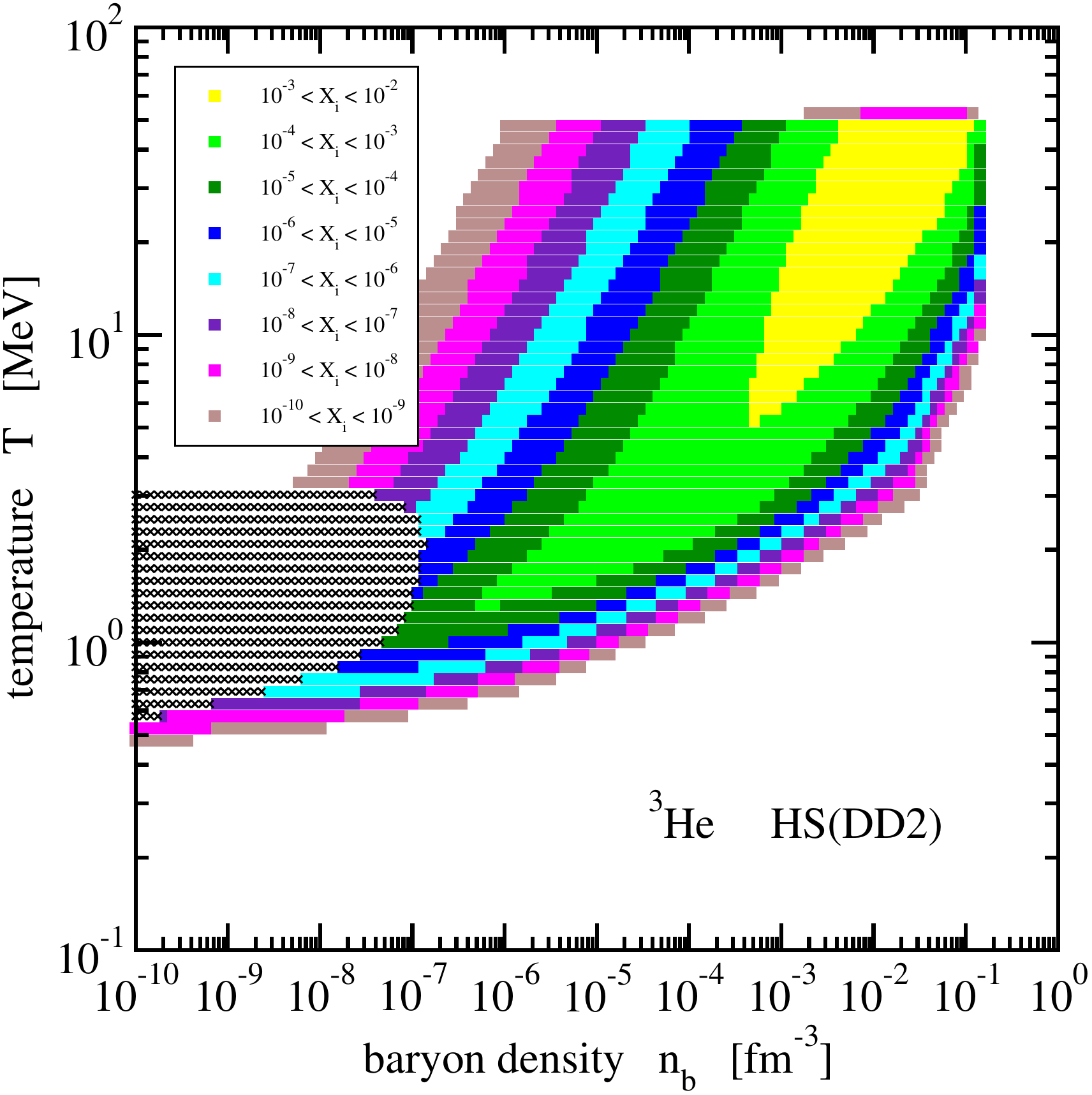}}
  }
  \caption{(Color online) Mass fraction $X_{t}$ of tritons ($^{3}$H)
    for the gRDF model (a) and the HS model (b) and
    mass fraction $X_{h}$ of helions ($^{3}$He)
    for the gRDF model (c) and the HS model (d)
    in neutron star matter.
    The area of datapoints outside the EoS table is indicated by crosses.}
  \label{fig:03}
\end{figure}

The observations for the deuteron and the $\alpha$ particle apply in a similar way
to the ${}^{3}$H and ${}^{3}$He nuclei, with isospin $1/2$ and 
opposite neutron-to-proton composition. A comparison in figure \ref{fig:03}
clearly shows that the more neutron-rich nucleus ${}^{3}$H is more likely to be
found in neutron star matter with low hadronic charge fraction $Y_{q}$, c.f.\
figure \ref{fig:01}, than the proton-rich nucleus ${}^{3}$He. In the gRDF model, 
both the triton and the helion are predicted to reappear at the highest temperatures
for baryon densities below approx.\ $10^{-6}$~fm${}^{-3}$. This can be explained by a simple
thermodynamic consequence since for $T > 70$~MeV binding energies have a rather minor
effect on the distribution functions. A similar feature is expected for the HS model,
if the cutoff for considering clusters was not introduced at $50$~MeV.

\begin{figure}[ht]
  \centerline{
    \mbox{}
    \hspace*{20pt}
    \subfigure[]{\includegraphics[width=0.52\textwidth]{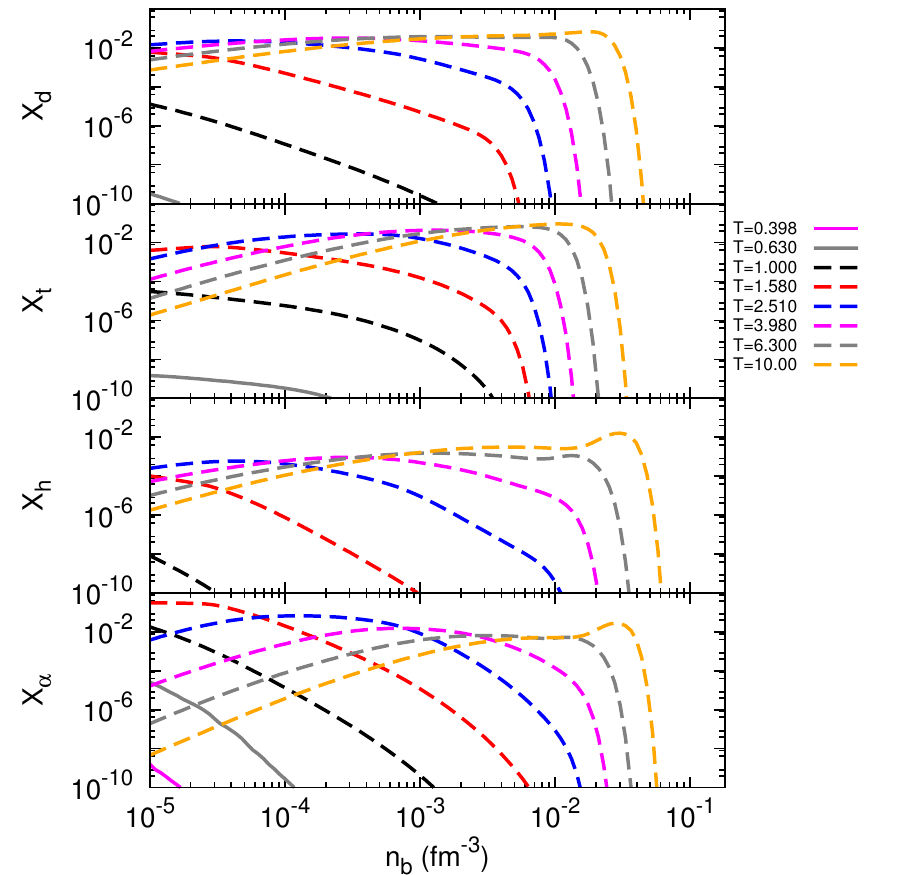}}
    \hspace*{4pt}
    \subfigure[]{\includegraphics[width=0.52\textwidth]{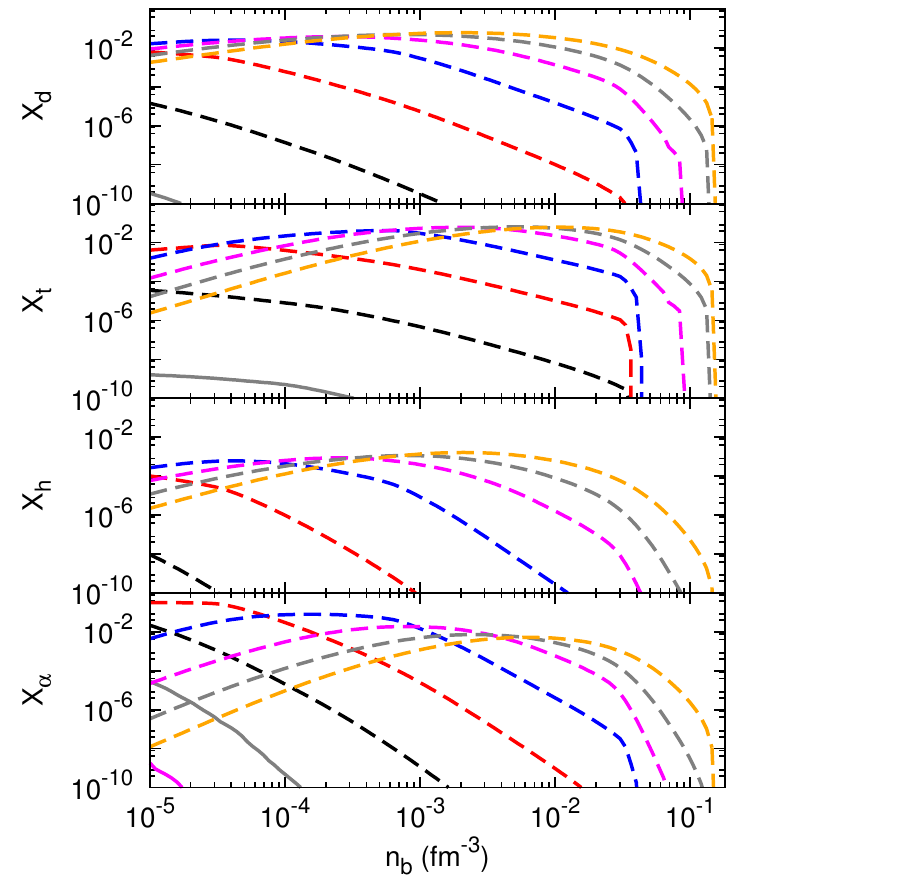}}
  }
  \caption{(Color online) Mass fractions of the light clusters, 
    $^2$H (d), $^3$H (t), $^3$He (h) and $^4$He ($\alpha$)
    for the gRDF model (a) and for the HS model (b) for several isotherms.}    
  \label{fig:04}
\end{figure}
\begin{figure}[ht]
  \centerline{
    \includegraphics[width=0.48\textwidth]{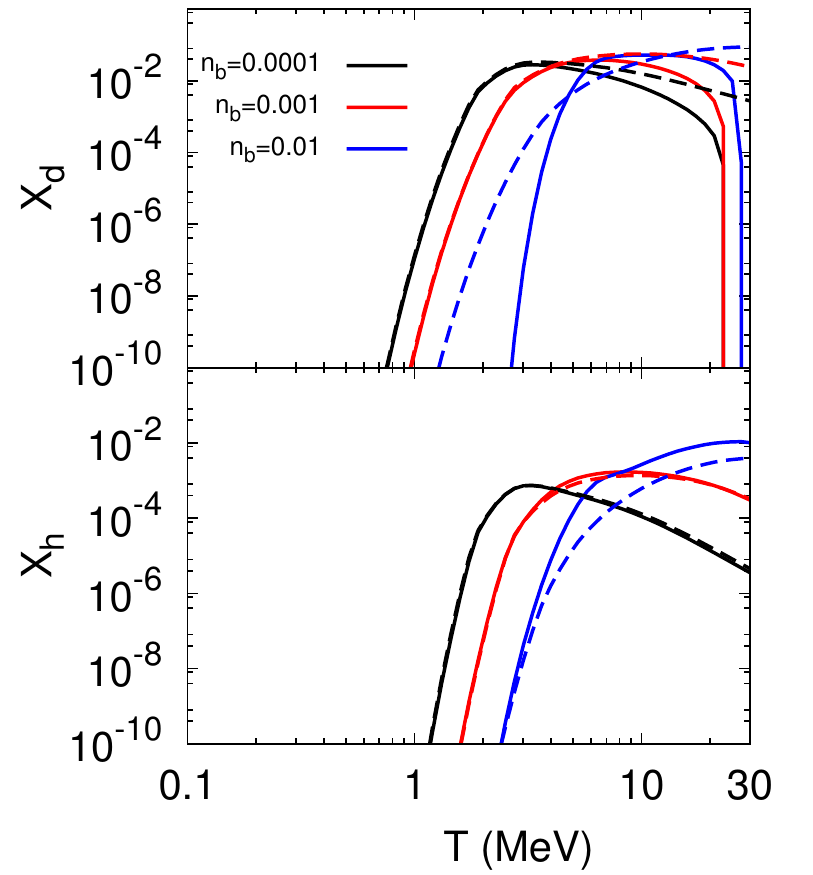}
    \hspace*{4pt}
    \includegraphics[width=0.48\textwidth]{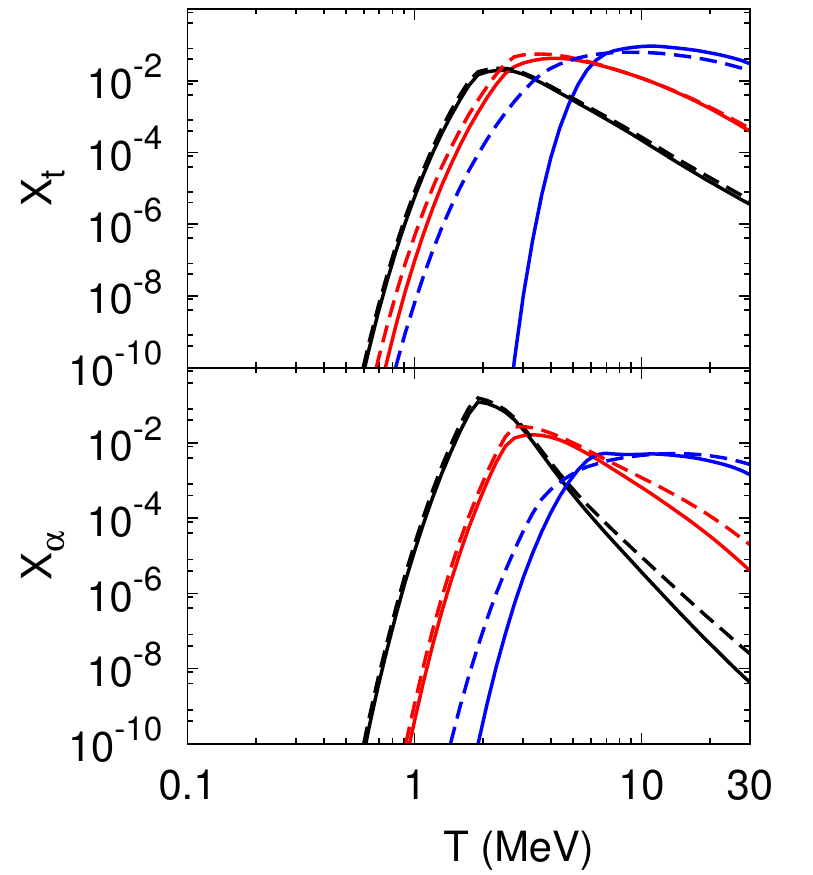}
  }
  \caption{(Color online) Mass fractions of the light clusters, $^2$H (d), $^3$H (t), $^3$He (h) and $^4$He ($\alpha$), for the gRDF model (full lines) and the HS model (dashed lines) as a function of the temperature for different values of the baryon number density.}    
  \label{fig:05}
\end{figure}

Larger differences in the predictions for light nuclei in the gRDF and HS model are visible
for densities above $10^{-3}$~fm${}^{-3}$. This becomes more apparent in figure
\ref{fig:04}, where the evolution of the mass fractions with density
is depicted for several isotherms
and a restricted density range, as compared to figures \ref{fig:02} and \ref{fig:03}. 
At densities lower than approx.\ $10^{-3}$~fm${}^{-3}$, the predictions
of the two models are almost identical. Here, mass-shifts or excluded-volume effects hardly affect
the mass fractions of light nuclei. But at higher densities, distinct differences appear.
The HS model with excluded-volume mechanism predicts a more gradual dissolution of the 
light clusters with density than the gRDF model with mass shifts.

At temperatures below approx.\ $0.5 - 1.0$~MeV, light clusters are hardly found
in neutron star matter. The variation of the mass fractions with temperature for
three fixed values of the baryon number density $n_{b}$ are shown in figure \ref{fig:05}.
At low densities of $10^{-4}$~fm$^{-3}$ or $10^{-3}$~fm$^{-3}$, there are only minor differences
between the gRDF and HS model in the mass fractions. Only at $10^{-2}$~fm${}^{-3}$ the predictions of the models differ more strongly. There are larger discrepancies at lower temperatures because here differences in the cluster masses or excluded-volume effects are more relevant in the thermodynamic description. In case of the gRDF model, the net deuteron fraction rapidly decreases with temperature, when $T$ becomes larger than approx.\ $20$~MeV. This is caused by the continuum contribution in the np(${}^{3}S_{1}$) channel that compensates the deuteron ground state contribution.

\begin{figure}[ht]
  \centerline{
    \subfigure[]{\includegraphics[width=0.48\textwidth]{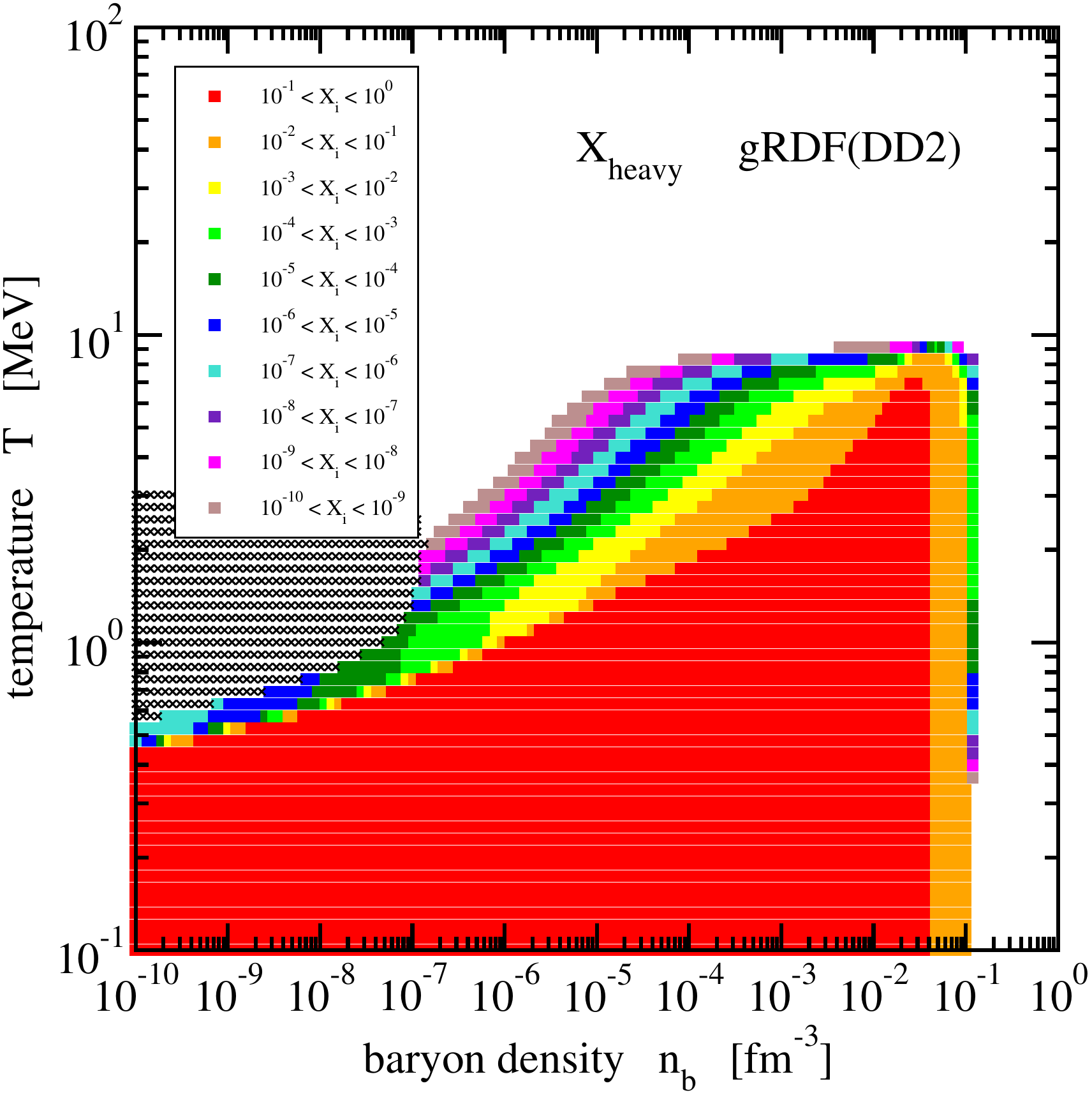}}
    \hspace*{4pt}
    \subfigure[]{\includegraphics[width=0.48\textwidth]{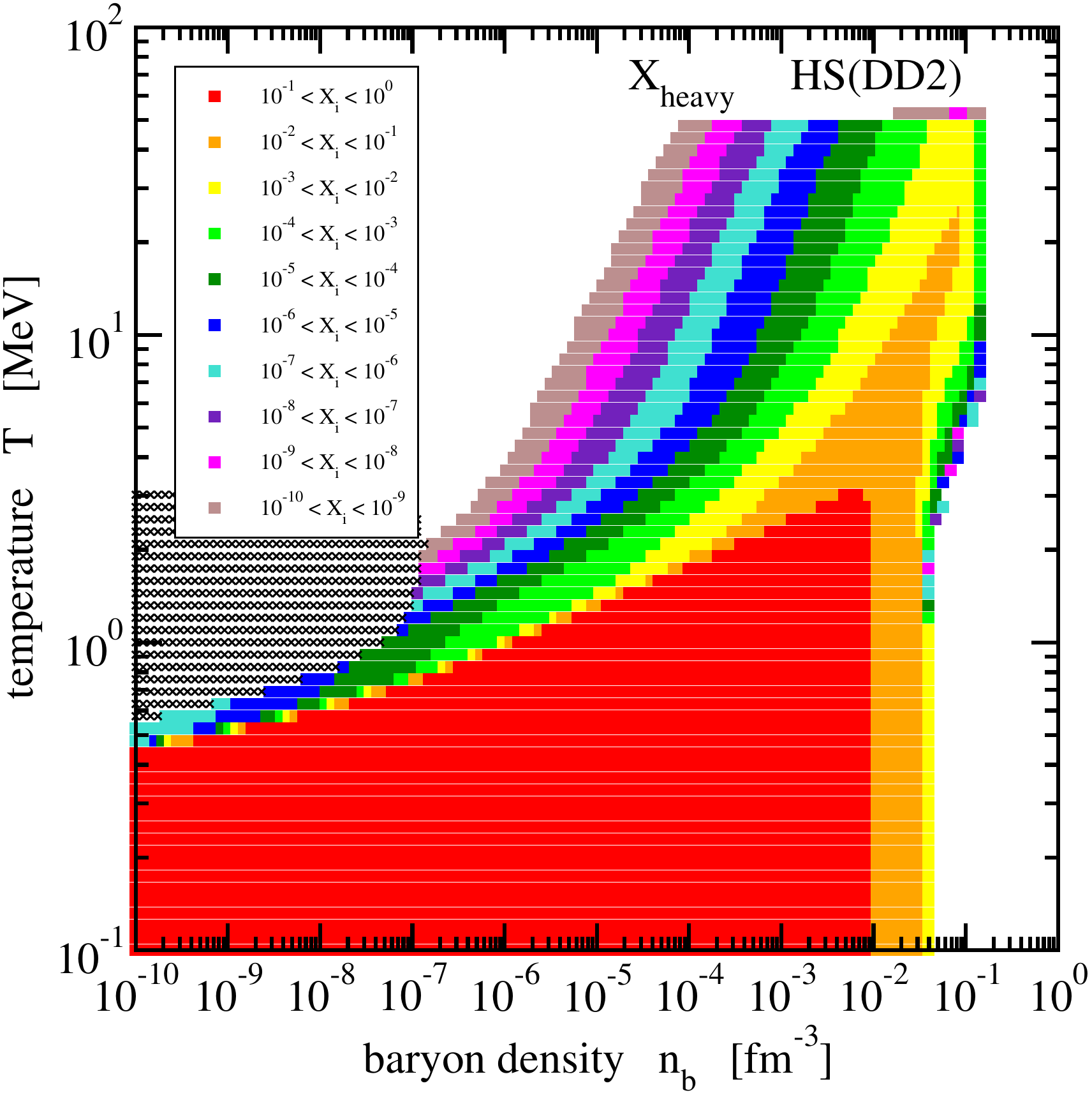}}
  }
  \centerline{
    \subfigure[]{\includegraphics[width=0.48\textwidth]{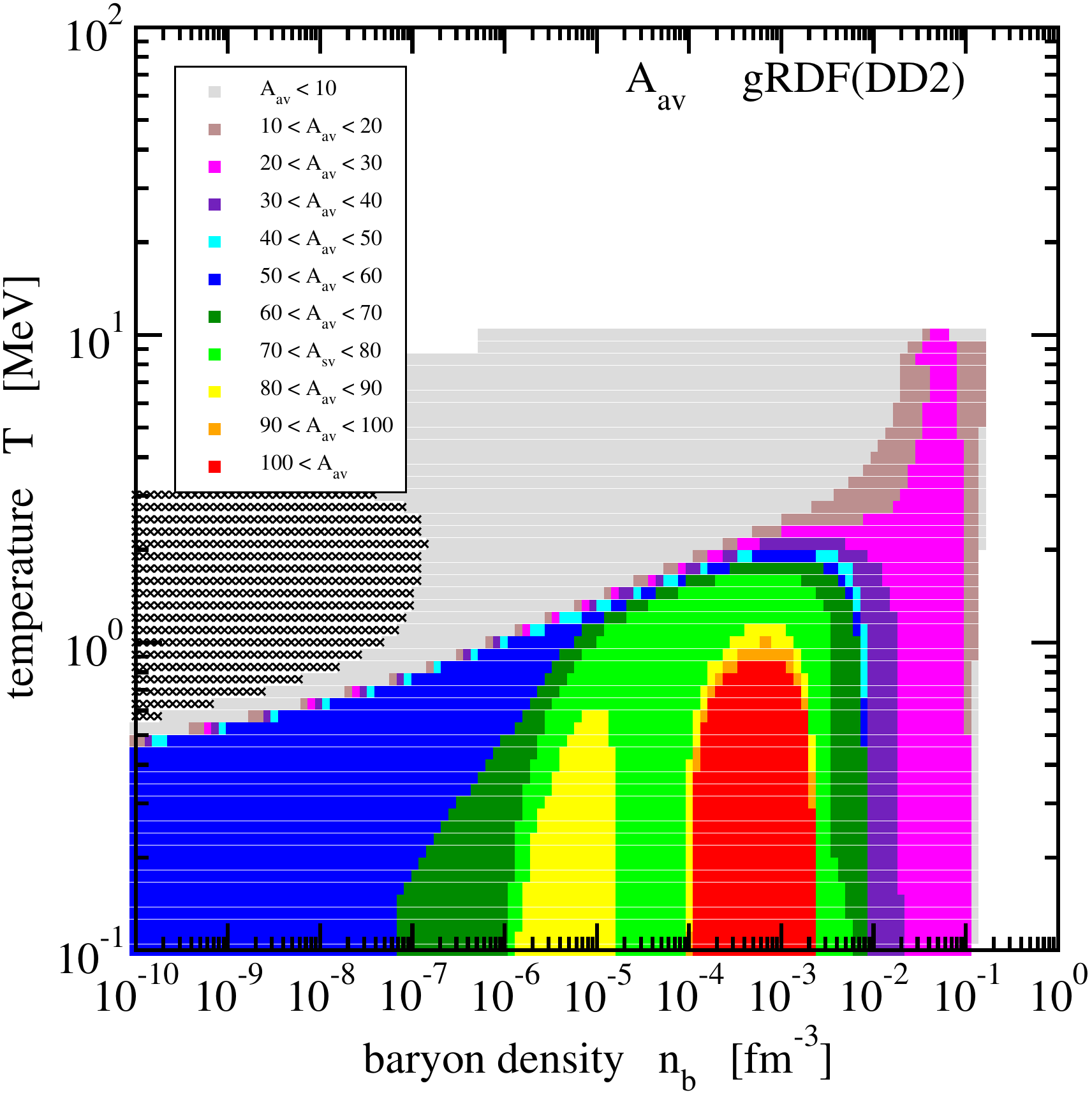}}
    \hspace*{4pt}
    \subfigure[]{\includegraphics[width=0.48\textwidth]{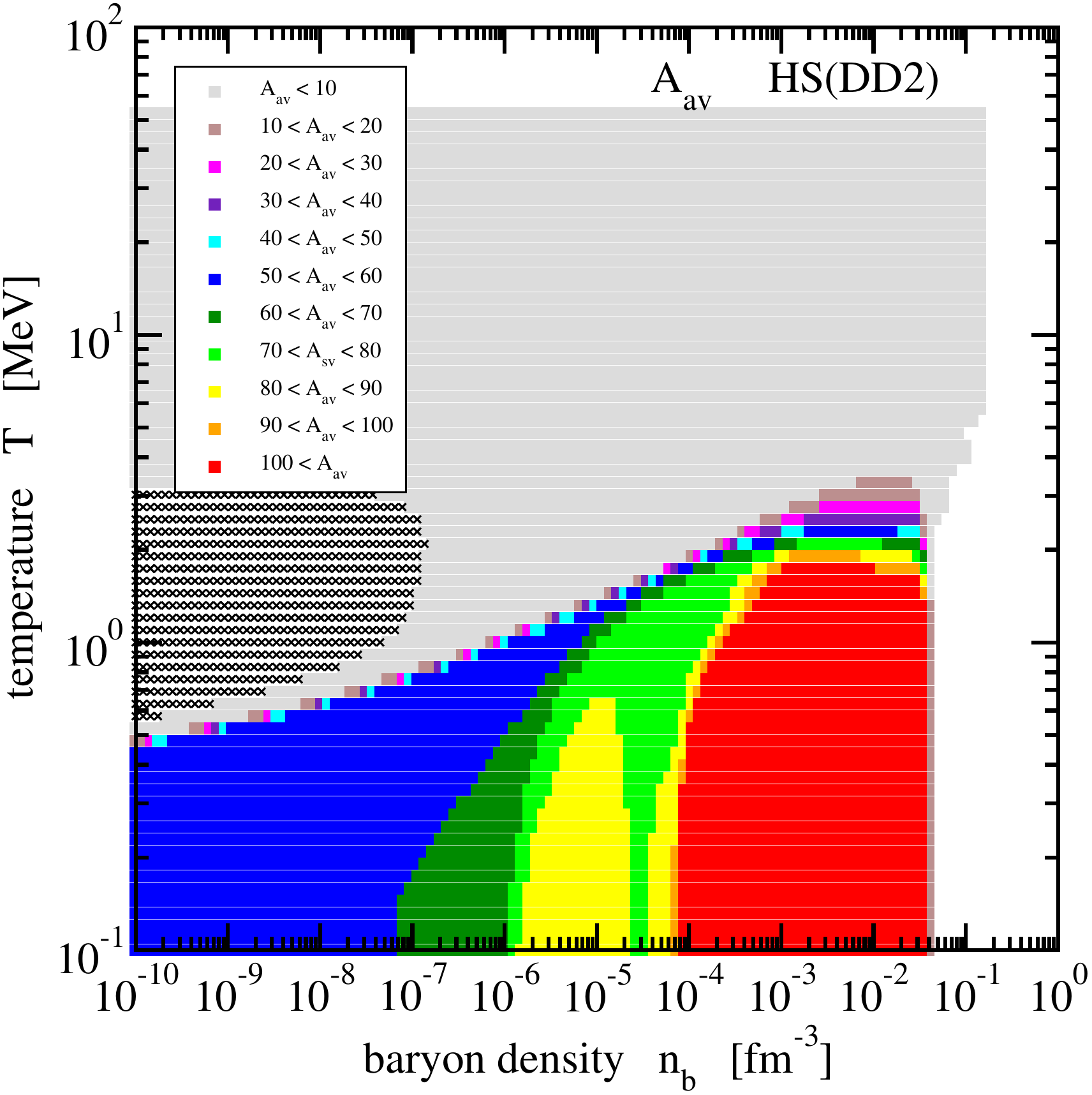}}
  }
  \caption{(Color online) Mass fraction $X_{\rm heavy}$ of heavy nuclei 
    for the gRDF model (a) and the HS model (b)
    and average mass number $A_{\rm av}$ of heavy nuclei
    for the gRDF model (c) and the HS model (d) in neutron star matter.
    The area of datapoints outside the EoS table is indicated by crosses.}
  \label{fig:06}
\end{figure}

The chemical composition of neutron star matter at low temperatures is dominated 
by heavy nuclei with mass numbers $A > 4$. This is evident from panels (a) and (b)
of figure \ref{fig:06}, that depict the total mass fraction of heavy nuclei
\begin{equation}
 X_{\rm heavy} = \sum_{(A,Z), A>4} X_{(A,Z)} 
\end{equation} 
with the mass fraction $X_{(A,Z)} = A n_{(A,Z)}$ of individual nuclei
as a function of
baryon density and temperature. Below $T \approx 2$~MeV, the gRDF model and the HS model
look very similar but the dissolution of heavy clusters occurs at lower densities
in the HS model, in the range from $10^{-2}$~fm${}^{-3}$ to $5 \cdot 10^{-2}$~fm${}^{-3}$,
whereas heavy nuclei survive up to densities somewhat above $10^{-1}$~fm${}^{-3}$.
The disappearance of heavy nuclei with increasing temperature for constant baryon density
shows some distinct differences, when the two models are compared. The $\gamma(T)$ factor
in the degeneracy factor (\ref{eq:g_i_T}) of the gRDF model causes the heavy clusters
to be removed from the system above approx.\ $10$~MeV. In the HS model, they appear
at even higher temperatures, since the excluded-volume mechanism as a geometric
concept is only depending on the density but not on the temperature. The reduction of
$X_{\rm heavy}$ in the HS model with increasing temperature 
is only a consequence of statistics. In panel (b) of figure \ref{fig:06} 
we notice again the artificial cluster cutoff at $50$~MeV in this model.

\begin{figure}[ht]
  \centerline{
    \subfigure[]{\includegraphics[width=0.48\textwidth]{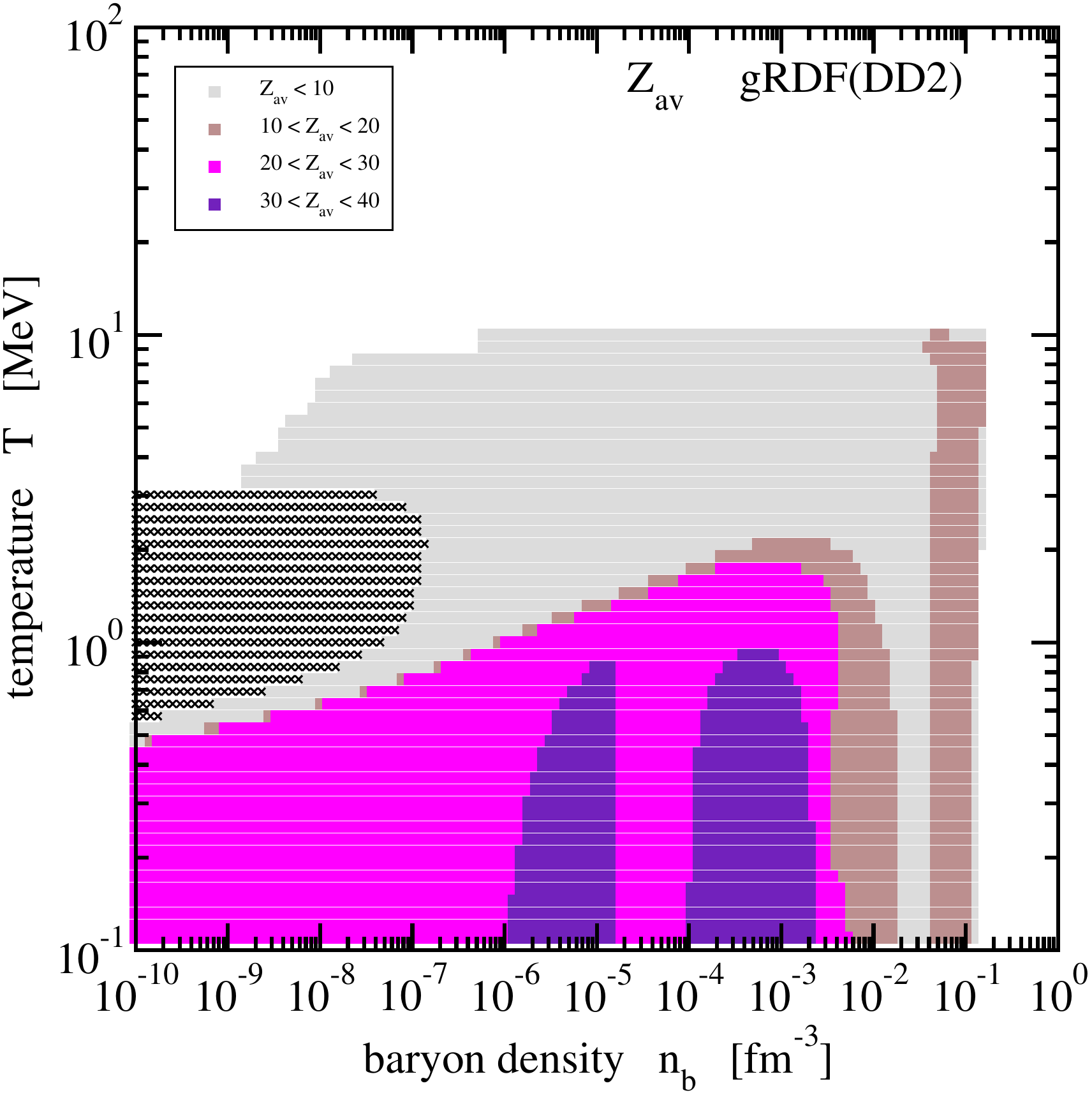}}
    \hspace*{4pt}
    \subfigure[]{\includegraphics[width=0.48\textwidth]{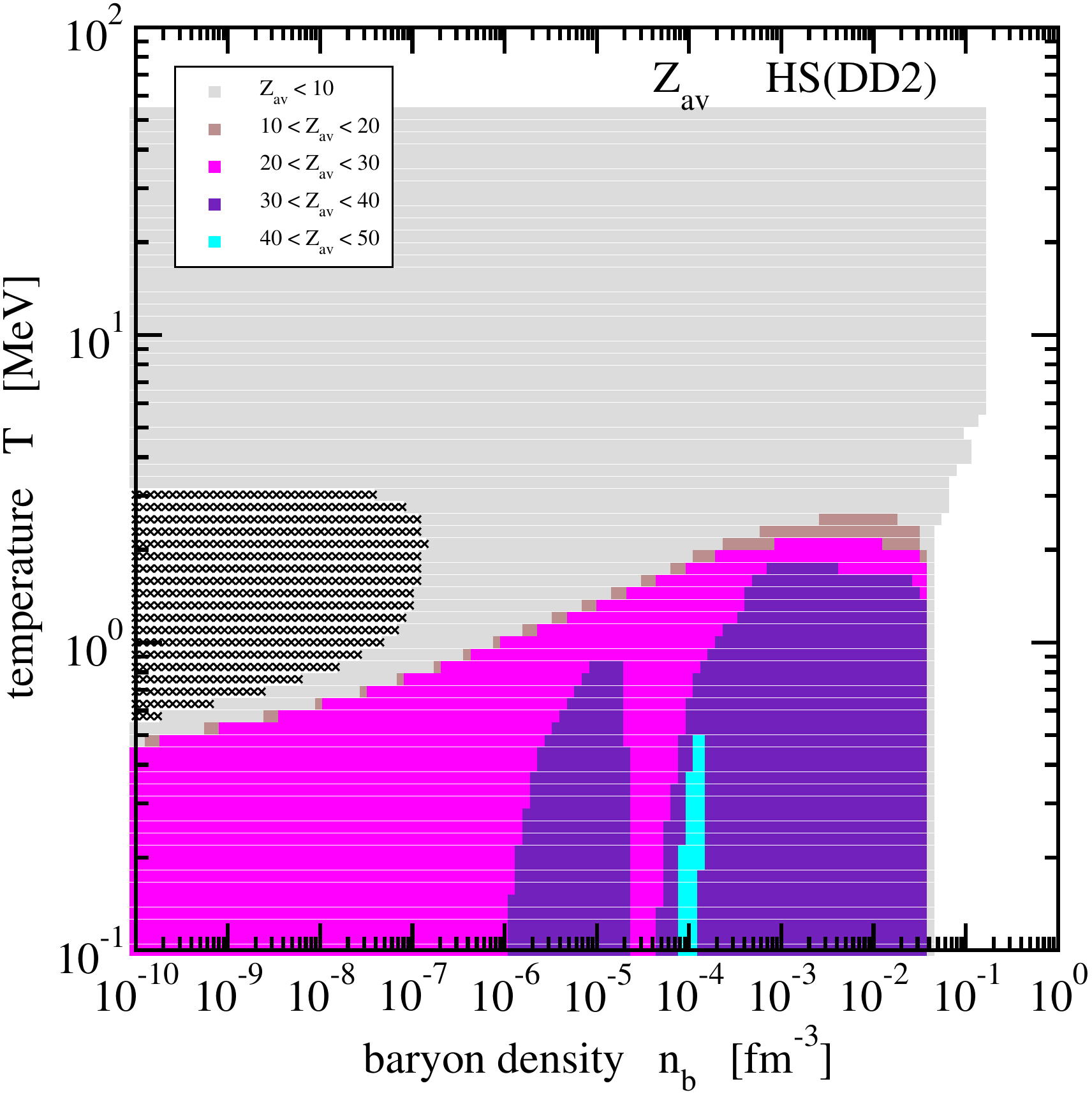}}
  }
  \centerline{
    \subfigure[]{\includegraphics[width=0.48\textwidth]{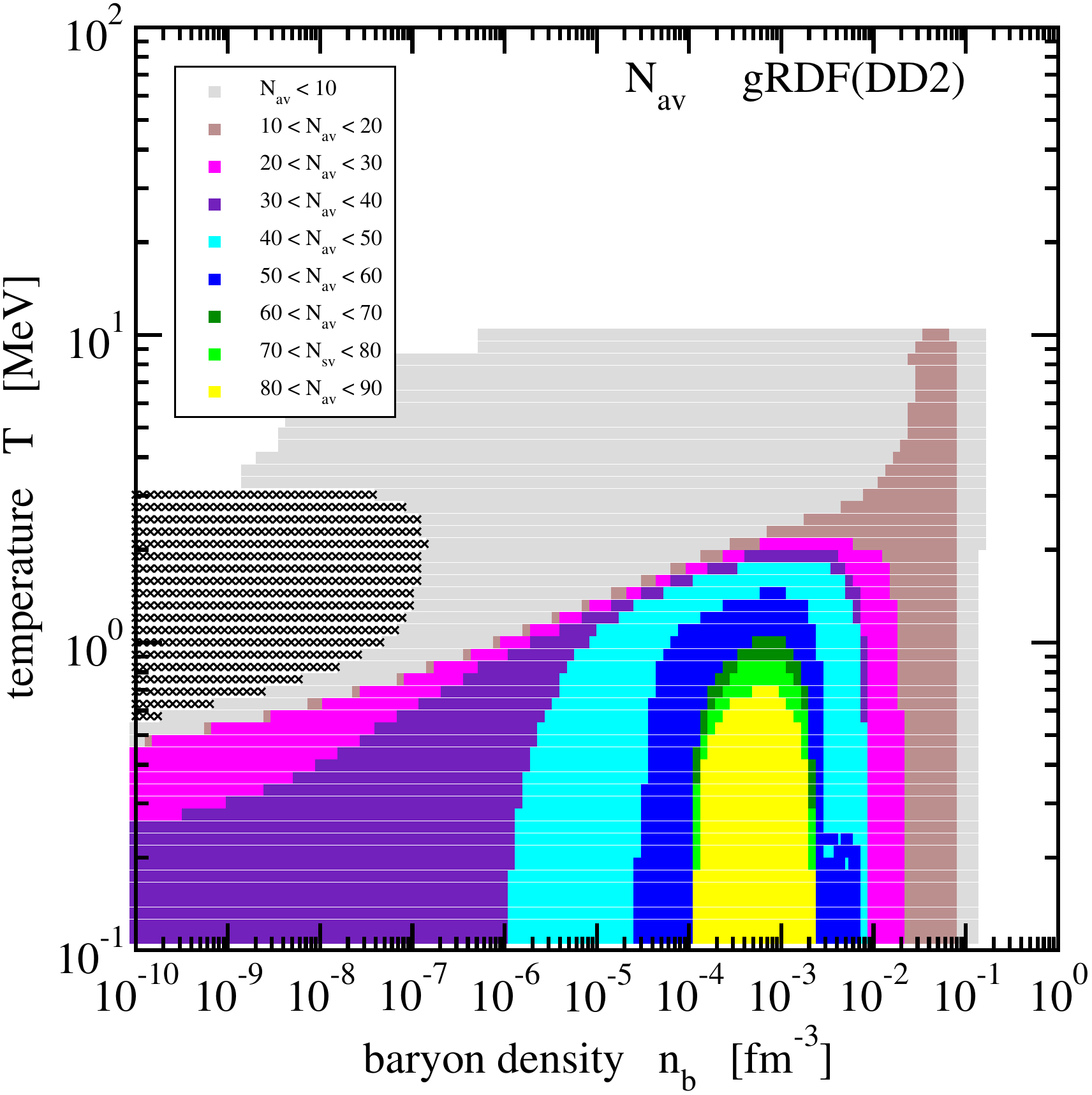}}
    \hspace*{4pt}
    \subfigure[]{\includegraphics[width=0.48\textwidth]{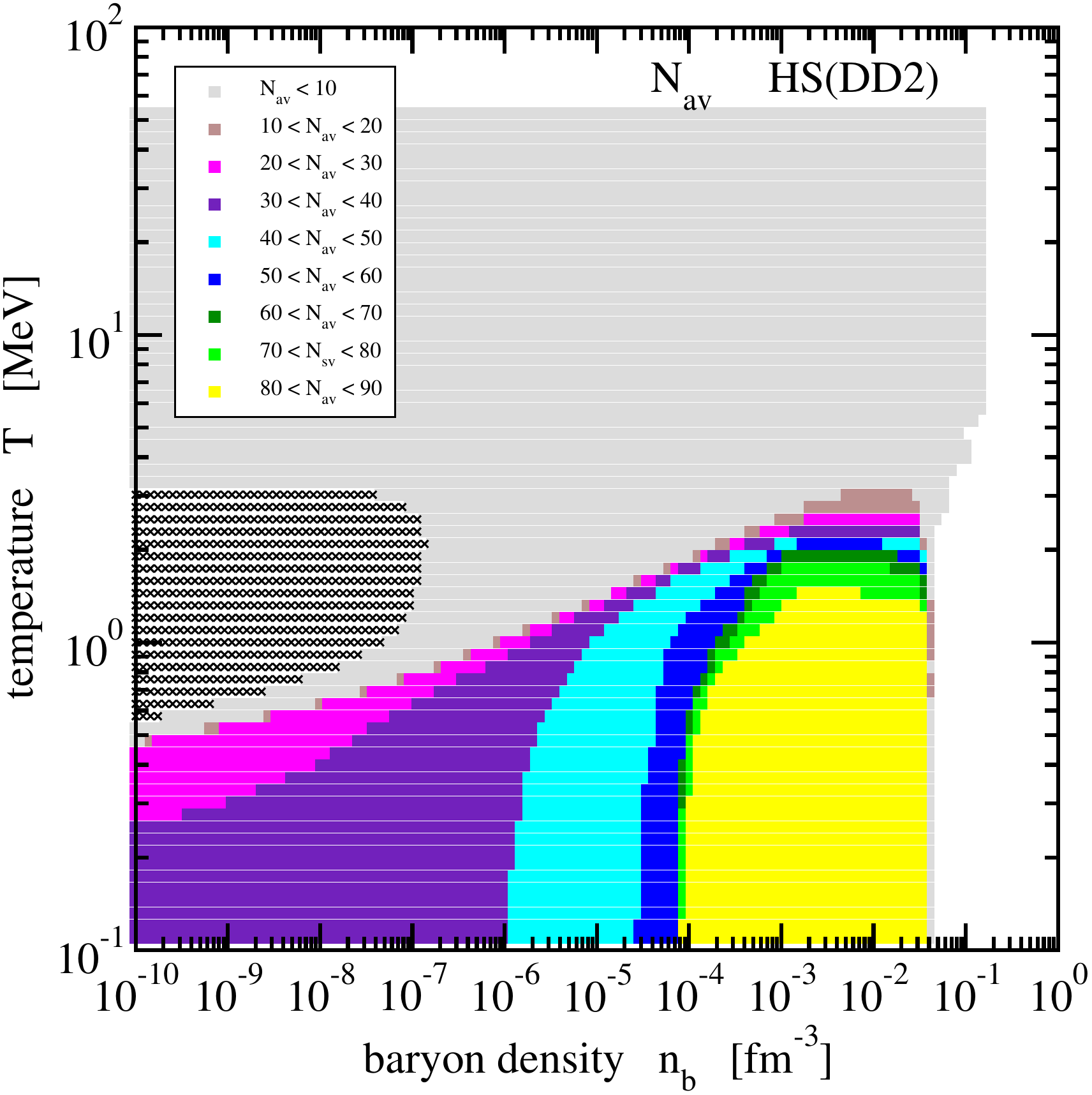}}
  }
  \caption{(Color online) Average charge number $Z_{\rm av}$ of heavy nuclei 
    for the gRDF model (a) and the HS model (b) and
    average neutron number $N_{\rm av}$ of heavy nuclei 
    for the gRDF model (c) and the HS model (d)
    in neutron star matter.
    The area of datapoints outside the EoS table is indicated by crosses.}
  \label{fig:07}
\end{figure}

In order to assess the importance of heavy nuclei in neutron star matter, 
not only the mass fraction $X_{\rm heavy}$, but also the size of the nuclei has to be
examined. In panels (c) and (d) of figure \ref{fig:06} the variation of the
average mass number
\begin{equation}
 A_{\rm av} = \frac{1}{X_{\rm heavy}} \sum_{(A,Z),A>4} A X_{(A,Z)}
\end{equation}
is illustrated, where the countour lines indicate a change by ten units.
For baryon densities below $10^{-5}$~fm${}^{-3}$, both models are very similar,
and the expected increase of the average mass number with density is observed for the lowest
temperatures. Above $10^{-5}$~fm${}^{-3}$, however, the models exhibit significant differences.
In the gRDF model, the highest average mass numbers are found for temperatures below $1$~MeV,
in the density range from $10^{-4}$~fm${}^{-3}$ to $2 \cdot 10^{-3}$~fm${}^{-3}$. 
At even higher densities, $A_{\rm av}$ continuously decreases to smaller values,
before the heavy clusters dissolve, cf.\ panel (a) of figure \ref{fig:06}. 
The situation is very different in the HS model. Here, the heaviest clusters 
survive almost until they disappear when the density increases. When the temperature
increases above $3$~MeV, the average mass number of heavy nuclei quickly shrinks, and the
chemical composition is governed by light clusters.

The change of the average mass numbers $A_{\rm av}$ is accompanied by a similar
variation of the average charge number $Z_{\rm av}$ and the average neutron number
$N_{\rm av}$ of heavy nuclei. These quantities are depicted in figure \ref{fig:07}
for the gRDF model and the HS model in the same style as $A_{\rm av}$ in figure
\ref{fig:06}. The difference between the models are again clearly visible. Both
$Z_{\rm av}$ and $N_{\rm av}$ decrease smoothly with increasing density above
$2 \cdot 10^{-3}$~fm${}^{-3}$ for constant temperature below $2$~MeV in the gRDF model,
and the maximum values are found in the density range 
from $10^{-4}$~fm${}^{-3}$ to $2 \cdot 10^{-3}$~fm${}^{-3}$. In the HS model,
large average charge and neutron number appear also at higher densities.
Because the matter is very neutron rich, cf.\ figure \ref{fig:01}, $N_{\rm av}$
reaches larger values than $Z_{\rm av}$. At the lowest temperatures, the sequence of
charge and neutron numbers with increasing baryon density is similar to that
expected in the crust of cold neutron stars.

\begin{figure}[t]
  \centerline{
    \mbox{}
    \hspace*{20pt}
    \subfigure[]{\includegraphics[width=0.5\textwidth]{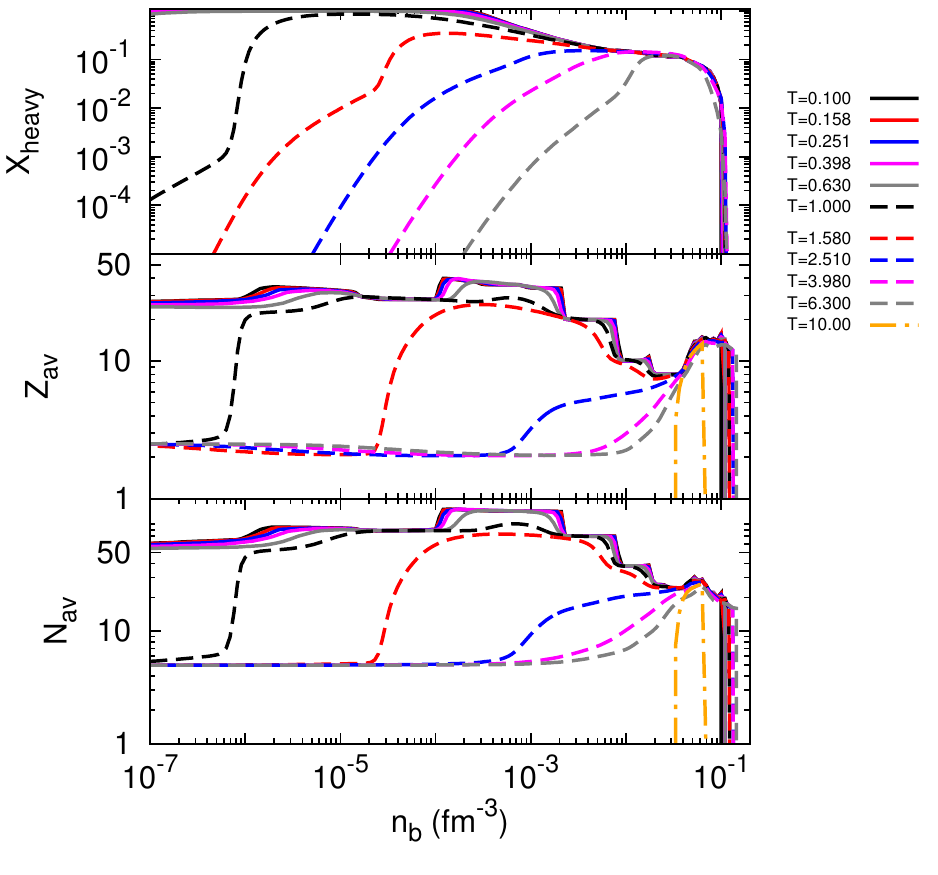}}
    \hspace*{4pt}
    \subfigure[]{\includegraphics[width=0.5\textwidth]{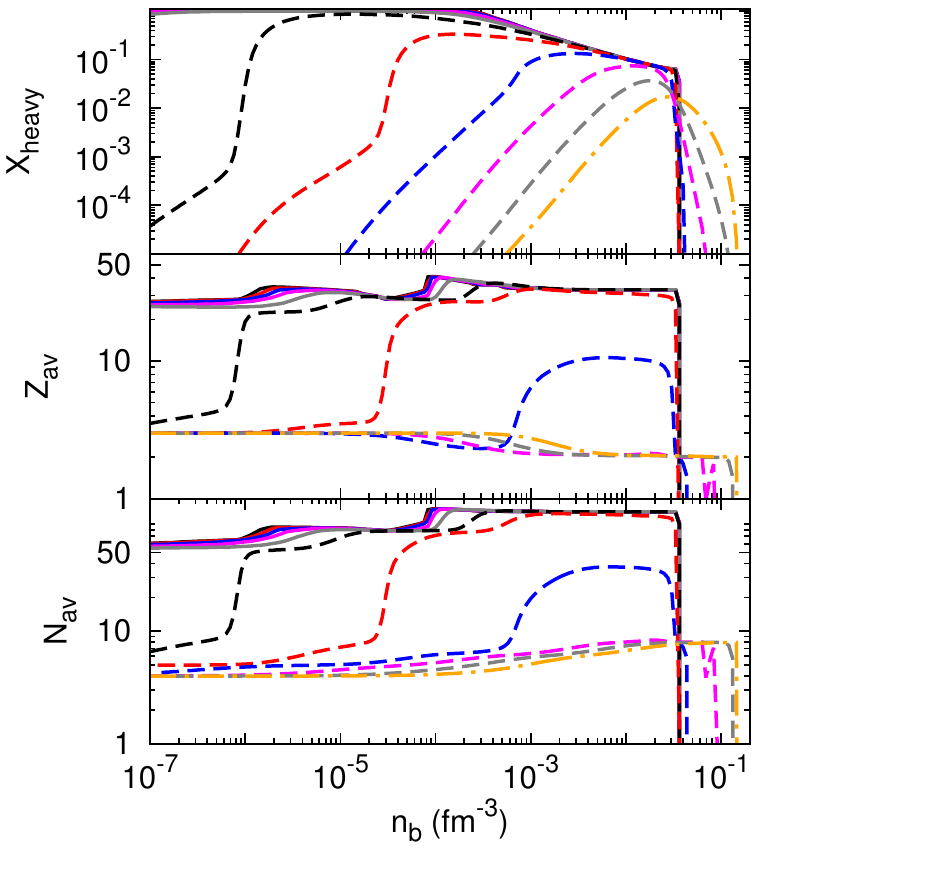}}
  }
  \caption{(Color online) Mass fraction $X_{\rm heavy}$, average charge number $Z_{\rm av}$ and average neutron number $N_{\rm av}$ of heavy nuclei as a function of the baryon number density $n_{b}$ for constant temperatures in the gRDF model (a) and in HS model (b).}
  \label{fig:08}
\end{figure}

The variation of $X_{\rm heavy}$, $Z_{\rm av}$ and $N_{\rm av}$ with increasing density for several isotherms in geometric progression is shown in figure~\ref{fig:08}, for a restricted range of densities.
Heavy nuclei dominate the chemical composition at temperatures below approx.\ $0.6$~MeV and baryon densities below approx.\ $2\cdot 10^{-4}$~fm${}^{-3}$, the neutron drip density in cold neutron star matter. 
In the HS model, heavy nuclei disappear abruptly at densities of about one fifth of the saturation density
at low temperatures, but survive for temperatures above $1$~MeV also at higher densities. 
They dissolve at almost the same density of $n_{\rm sat}/3$, irrespective of the temperature in the gRDF model. Shell effects in the average charge and neutron numbers are clearly visible
at low temperatures since the distribution of heavy nuclei is dominated by a few species.
The jumps at the magic numbers are washed out with increasing temperature. A comparison of the gRDF results with those of the HS model shows that the size of the heavy clusters gradually reduces with
increasing density in the former model but stays almost constant up to the dissolution density in the latter model.

\begin{figure}[t]
  \centerline{
    \includegraphics[width=1.\textwidth]{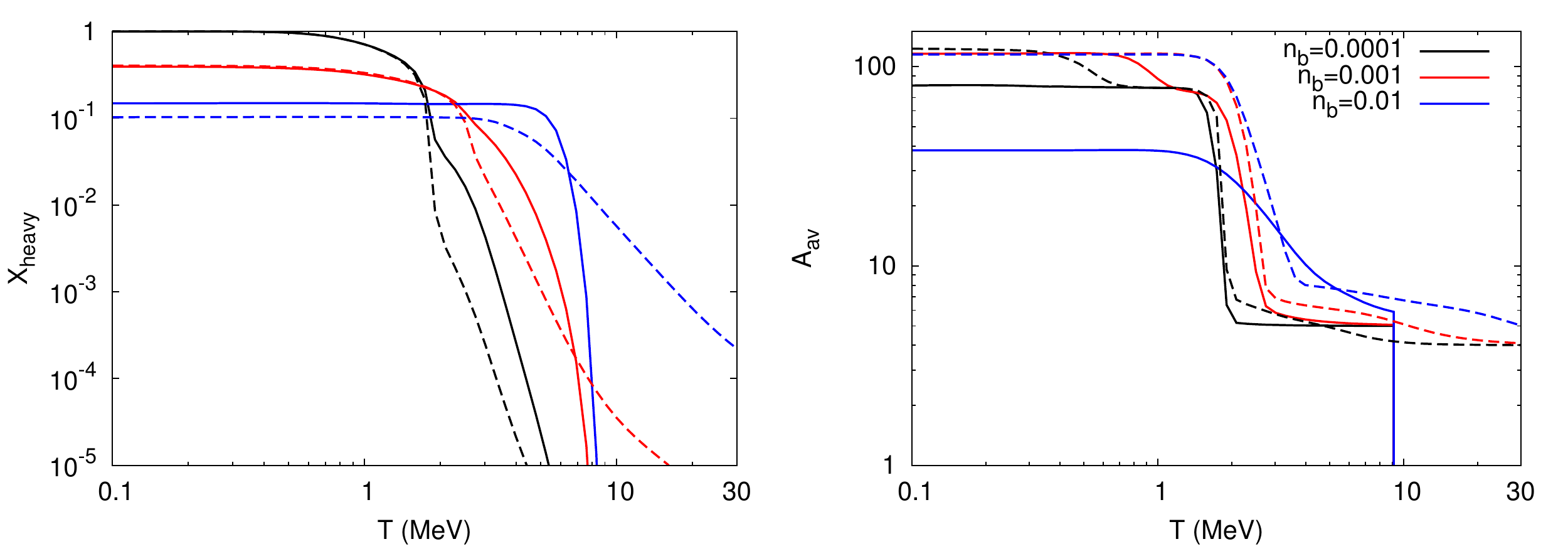}
  }
  \caption{(Color online) Mass fraction $X_{\rm heavy}$ and average mass number $A_{\rm av}$ of heavy nuclei as a function of the temperature $T$ for constant baryon number densities $n_{b}$ in the gRDF model (full lines) and in the HS model (dashed lines).}
  \label{fig:09}
\end{figure}

The fraction of heavy nuclei and their average mass number as a function of the temperature
for fixed baryon densities is depicted in figure~\ref{fig:09}. The dissolution of heavy nuclei
with increasing temperature is evident with a steeper decrease in the gRDF model, where heavy nuclei
are suppressed, if $T$ approaches $10$~MeV. At the highest density of $10^{-2}$~fm${}^{-3}$, differences in $X_{\rm heavy}$ between the models, even at low temperatures, are observed. The evolution of the average mass fraction with temperature also exhibits clear differences, depending on the chosen value of $n_{b}$ in accordance with the lower panel of figure~\ref{fig:06}. The sudden change of $A_{\rm av}$ with the temperature at low $T$ is again related to shell effects that are taken care of in both models due to the use of realistic mass tables.

\subsection{Thermodynamic properties}

\begin{figure}[ht]
  \centerline{
    \subfigure[]{\includegraphics[width=0.48\textwidth]{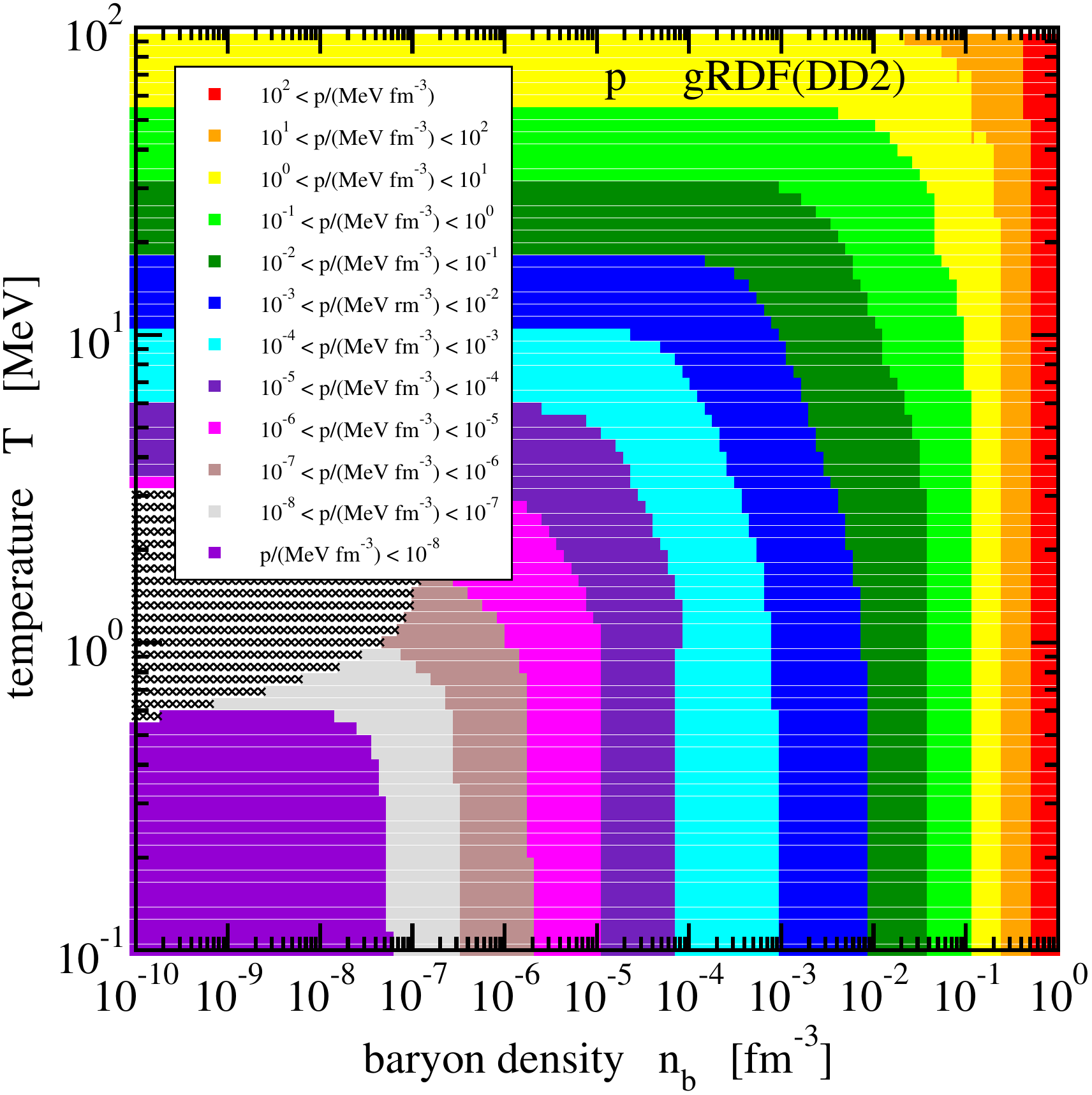}}
    \hspace*{4pt}
    \subfigure[]{\includegraphics[width=0.48\textwidth]{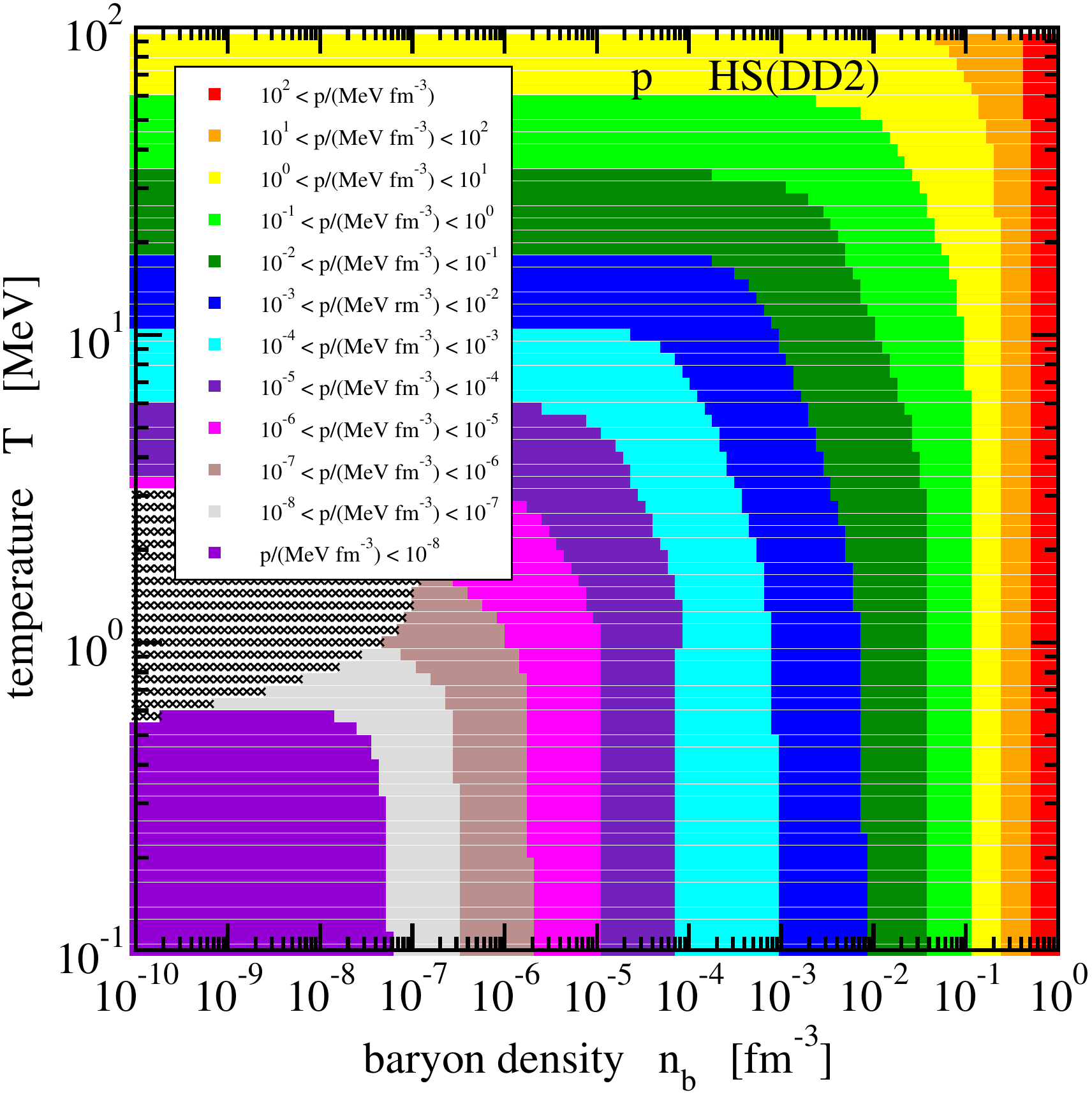}}
  }
  \centerline{
    \subfigure[]{\includegraphics[width=0.48\textwidth]{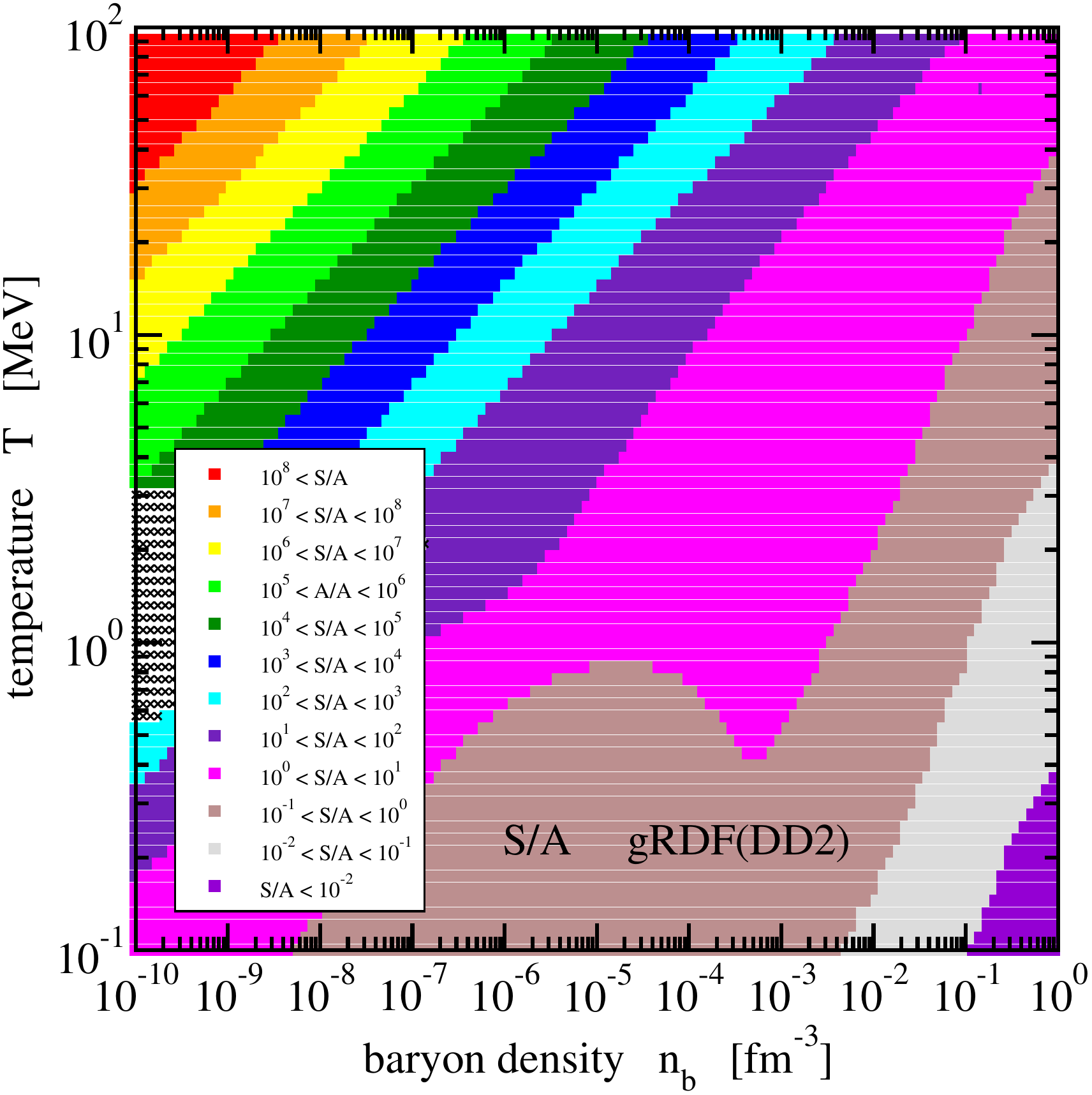}}
    \hspace*{4pt}
    \subfigure[]{\includegraphics[width=0.48\textwidth]{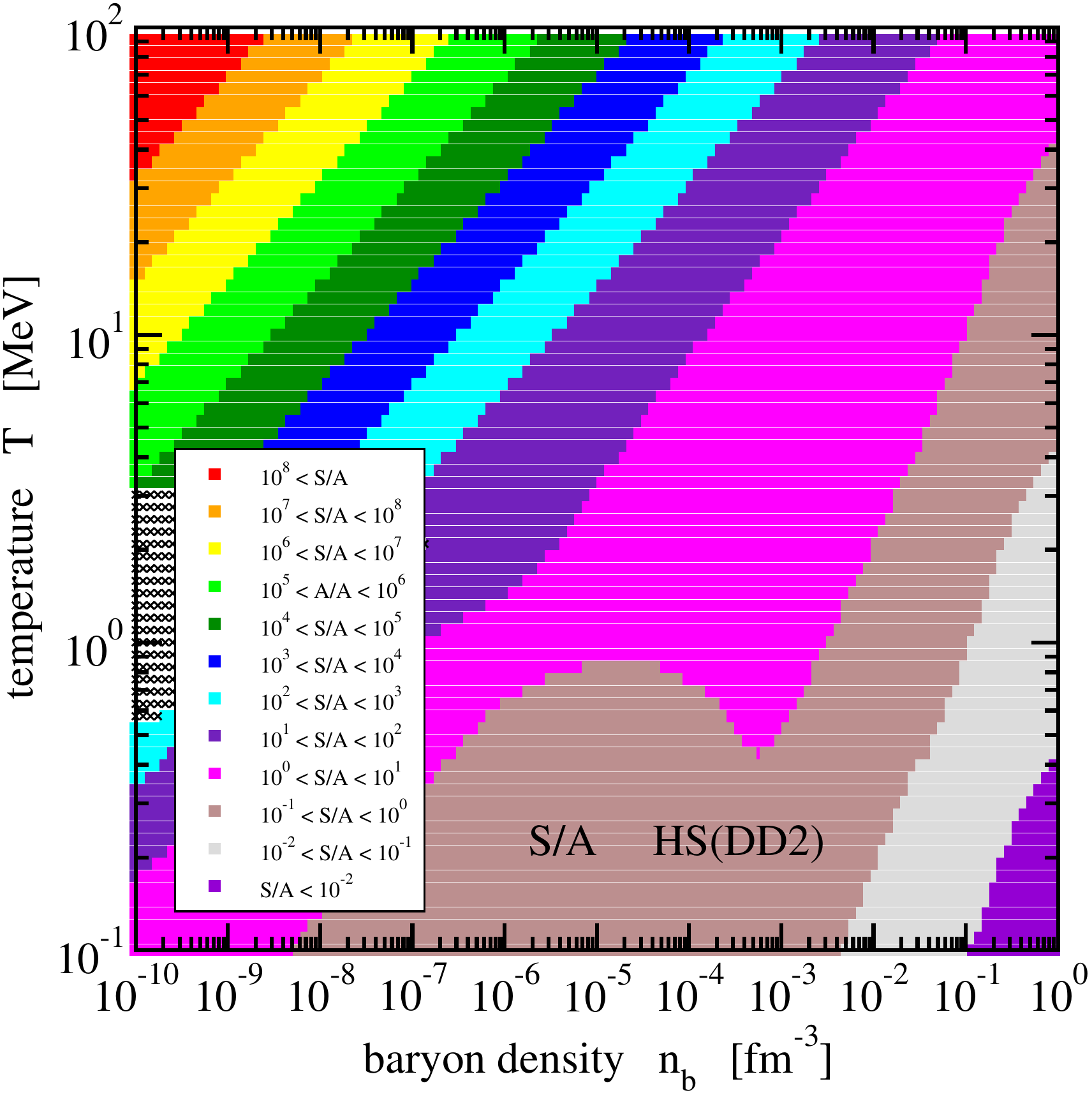}}
  }
  \caption{(Color online) Pressure $p$
    for the gRDF model (a) and the HS model (b) and
    entropy per baryon $S/A$
    for the gRDF model (c) and the HS model (d)
    in neutron star matter.
    The area of datapoints outside the EoS table is indicated by crosses.}
  \label{fig:10}
\end{figure}

Besides the chemical composition of neutron star matter, the predictions of the gRDF model
and the HS model for thermodynamic quantities can be compared. As representative examples,
the pressure $p$ and the entropy per baryon $S/A = s/n_{b}$ are depicted
in figure \ref{fig:10} as a function of baryon density and temperature. The pressure and
entropy per baryon along lines that separate differently colored regions differ by a factor of ten.
There is a systematic increase of the pressure with $n_{b}$ and $T$ reaching the largest values
at baryon densities approaching $1$~fm${}^{-3}$. In contrast, the highest entropy per baryon
is found at the highest temperatures close to $100$~MeV, at the lowest baryon densities. 
In general, $S/A$ decreases with decreasing temperature and increasing baryon density.
There is an exception from this trend in the region close to the line of constant $S/A=1$,
where the chemical composition of the matter rapidly changes. 
Overall there is a surprising agreement of the pressure and entropy per baryon between the two models, despite the very different approaches to model the dissolution of nuclei. A similar concordance of thermodynamic quantities,
such as the free energy per baryon or chemical potentials is observed. Hence, we refrain
from showing the corresponding figures. 

\section{Conclusions}
\label{sec:concl}

In the present work, we presented and compared two different theoretical approaches to model the equation of state and in particular the formation and dissolution of nuclear clusters in stellar matter: a statistical model with excluded-volume effects (HS model) and a generalized relativistic density functional (gRDF model) with an in-medium change of the cluster masses. The theoretical formalism of the two approaches was presented in detail and the essential differences were delineated. Both models use a relativistic mean-field description for the nucleonic part with the same parametrisation of the density dependent meson-nucleon couplings, but they differ in the treatment of the cluster degrees of freedom. Data on the chemical composition and on thermodynamic properties are available in tabular form for both models in a wide range of baryon densities, temperatures and isospin asymmetries in the CompOSE format. The corresponding tables are taken from the CompOSE website.

We were interested mainly in the evolution of the light and heavy clusters with density and temperature. For that purpose, we studied neutron star matter, i.e.\ charge-neutral matter in $\beta$ equilibrium, and observed that the gRDF and the HS models predict some differences for the mass fractions and average sizes of the  clusters. Both models behave similarly for low temperatures and low densities where mass shifts or excluded-volume effects are negligible. When the baryon density approaches the nuclear saturation density, the dissolution of clusters is more gradual in the gRDF model than in the HS model. Heavy clusters disappear rather abruptly with increasing density in the HS model, whereas in the gRDF model, their size gradually reduces until they melt.

Above temperatures of 50 MeV,  light and heavy clusters are artificially suppressed in the HS model. In the gRDF model, heavy  clusters are removed from the system when the temperature exceeds approximately $10$~MeV due to the specific temperature dependence of the degeneracy factors. In the HS model, they do appear at even higher temperatures, since the excluded-volume mechanism does not depend on the temperature, but only on the density. 
 
Despite the differences in the cluster abundances and properties, both models show overall a very good agreement of the thermodynamic properties. Thus it can be expected that dynamical simulations of core-collapse supernovae or neutron star mergers are only marginally affected by the choice of the cluster description in the equation of state as long as the interaction model for the nucleonic part is identical. However, the chemical composition of stellar matter could affect processes such as the neutrino transport or nucleosynthesis in the above mentioned astrophysical scenarios. These consequences of the theoretical cluster description can be explored in future studies.

\section*{Acknowledgments}

This work was supported by NewCompStar, COST Action MP1304, 
by the Helmholtz Association (HGF) through the
Nuclear Astrophysics Virtual Institute (VH-VI-417), and by the DFG through
grant No.\ SFB1245. H.P. is supported by FCT under Project No. SFRH/BPD/95566/2013. She is very thankful to S.T. and the Theory group at GSI for the kind hospitality during her stay there.

\bibliographystyle{ws-rv-van}
\bibliography{paper}


\end{document}